\documentclass[12pt]{article}
\usepackage[dvips]{graphicx}
\usepackage{latexsym}
\usepackage[abs]{overpic}
\usepackage{wrapfig}
\usepackage{amssymb}
\usepackage{amsmath}
\usepackage{bm}

\input epsf
\newcommand{\R}{R}

\newcommand{\skyp}[1]{}

\def\Z {\bb{Z}}
\def\R {\bb{R}}

\def\rem#1{}

\def\C{{\mathbb{C}}} 
\def\D{{D^2}}

\def\N{{\mathbb{N}}}

\def\R{{\mathbb{R}}}
\def\Z{{\mathbb{Z}}} 

\def\th{{\theta}}
\def\mth{{\vartheta}}
\def\vphi{{\varphi}}

\def\ep{{\epsilon}}
\def\bep{{\bar{\epsilon}}}
\def\vep{{\varepsilon}}
\def\bpsi{{\bar{\psi}}}
\def\bphi{{\bar{\phi}}}
\def\bF{{\bar{F}}}
\def\lam{{\lambda}}
\def\blam{{\bar{\lambda}}}

\def\mN{\mathcal{N}}

\def\sig{{\sigma}}
\def\lam{{\lambda}}

\def\del{{\partial}}

\newcommand\non{\nonumber \\}
\newcommand{\bel}{\begin{eqnarray}}
\newcommand{\ee}{\end{eqnarray}}

\def\del{{\partial}}

 \newcommand{\todayd}{\the\year/\the\month/\the\day}
 \newcommand{\todaye}{\the\year--\the\month--\the\day}

\makeatletter
\@addtoreset{equation}{section}
\makeatother
\begin{document}

\begin{titlepage}

\bigskip
\hfill\vbox{\baselineskip12pt
\hbox{KIAS-P14056}
}
\bigskip\bigskip\bigskip\bigskip

\begin{center}
\large{\bf Localization of 3d $\mathcal{N}=2$ Supersymmetric Theories on $S^1 \times \D$ }
\end{center}

\bigskip
\bigskip
\bigskip
\bigskip
\centerline{ \large Yutaka Yoshida$^{\heartsuit}$ and  Katsuyuki Sugiyama$^{\spadesuit}$}
\bigskip
\bigskip
\centerline{$^{\heartsuit}$\it School of Physics, Korea Institute for Advanced Study (KIAS)}
\centerline{\it  85 Hoegiro Dongdaemun-gu, Seoul, 130-722,  Korea }
\centerline{ yyyyosidaATgmail.com}
\bigskip
\bigskip
\centerline{$^{\spadesuit}$\it Department of Physics, Graduate School of Science, Kyoto University}
\centerline{\it  Kitashirakawa Oiwake-cho, Sakyoku, Kyoto 606-8502, Japan \ \ }
\centerline{ sugiyamaATscphys.kyoto-u.ac.jp}
\begin{abstract}
We study  three dimensional $\mathcal{N}=2$ supersymmetric Chern-Simons-Matter theories on 
the direct product of a circle and a two dimensional hemisphere ($S^1 \times \D$) with specified boundary conditions by the method of localization.
We construct boundary interactions to cancel the supersymmetric variation of the three dimensional superpotential term and the Chern-Simons term 
and show inflows of the bulk-boundary anomalies. It finds that 
the boundary  conditions  induce two dimensional $\mathcal{N}=(0,2)$ type supersymmetry on the boundary torus. 
We also study the relation between the 3d-2d coupled  partition function of our model and three dimensional holomorphic blocks. 

\end{abstract}

\end{titlepage}

\newpage
\baselineskip=18pt

\tableofcontents
\section{Introduction}

After the prominent work by Pestun \cite{Pestun:2007rz}, various supersymmetric gauge theories on curved spacetimes have been investigated  and 
rigid supersymmetries on these backgrounds have been constructed. By evaluating fixed point sets of supersymmetric transformations combined with 
the method of localization, one can calculate exact partition functions of these supersymmetric theories.
For example, a round sphere $S^3$ \cite{Kapustin:2009kz, Jafferis:2010un, Hama:2010av} and an ellipsoid $S^3_b$ \cite{Hama:2011ea}
were investigated as suitable backgrounds of three dimensional (3d) supersymmetric theories.
The superconformal indices on $S^1 \times S^2$ were studied in \cite{Kim:2009wb, Imamura:2011su},  which 
 count the  number of BPS operators.
These functions play important roles in studying   M5-branes, 
IR dualities among supersymmetric theories and the AdS/CFT correspondence.

It is widely believed that these supersymmetric theories have interesting properties of  the factorizations with 
fundamental building blocks. 
For example, in three dimensional cases, it was conjectured in \cite{Beem:2012mb} that
 $\mathcal{N}=2$ supersymmetric partition functions on $S^3_b$ and the superconformal indices on $S^1 \times S^2$ 
can be expressed as bilinear forms of  two identical building blocks $\mathcal{B}^{\alpha} (x, q)$:
\bel
&&Z_{S^3_b \,  \text{or} \,  S^1 \times S^2}= \sum_{\alpha}  \mathcal{B}^{\alpha} (x, q)  \mathcal{B}^{\alpha} (\tilde{x}, \tilde{q}).
\ee
The authors of \cite{Beem:2012mb} have proposed some general rule to write down these  
universal blocks $\mathcal{B}^{\alpha} (x, q)$ called {\em holomorphic blocks}. 
By the  analogy of the  pure Chern-Simon theory \cite{Witten:1988hf} 
and the topological - anti-topological fusion in two dimensions \cite{Cecotti:1991me}, it is expected that
 holomorphic blocks should be  partition functions of three dimensional Chern-Simons-Matter theories on a solid torus (Melvin cigar). 
In order to  clarify the relation between their partition functions (indices) and holomorphic blocks, 
it is  interesting  to construct explicitly  $\mathcal{N}=2$ supersymmetric Chern-Simons-Matter theories on a solid torus.
When the spacetime has boundaries, one may  introduce physical  degrees of freedom on the boundaries and  classify what types of BPS boundaries are allowed in the supersymmetric Chern-Simons-Matter theories .

For example, half BPS boundary conditions have been studied in $\mN=1$ supersymmetric pure Chern-Simons theory  \cite{Sakai:1989nh} and 
  supersymmetric   Chern-Simons-Matter theories  \cite{Berman:2009kj,Faizal:2011cd}.  
 Theories on the boundary  studied so far are realized as super Wess-Zumino-Witten (WZW)  models and it is  natural 
to ask whether other types of boundary interactions exist or not ? 
For example, in two dimensional $\mathcal{N}=(2,2)$ theories, the boundary interaction is described in terms of the matrix factorization
so that the effect of the supersymmetric variation of the superpotential \cite{Kapustin:2002bi, Herbst:2008jq} is canceled. 
Higher dimensional analogues of the matrix factorization have not been much studied yet.

In this article, we investigate  three dimensional $\mathcal{N}=2$ supersymmetric Chern-Simons-Matter theories on 
the direct product of a circle and a two dimensional hemisphere ($S^1 \times \D$) toward the understanding of properties of the building blocks 
and the factorization. We impose suitable boundary conditions for 
$\mathcal{N}=2$ multiplets consistent with the supersymmetric transformations. Under the boundary conditions, the
super Yang-Mills action and the kinetic action of the chiral multiplet are written as $Q$-exact forms without surface terms and 
these actions are invariant under the supersymmetric transformations. On the other hand, in the presence of the boundary,
 the Chern-Simons term is neither  invariant under the supersymmetric transformation nor the gauge transformation. Thus 
we have to introduce some boundary term to compensate these two variations. 
Here we propose two possible ways to make the theory gauge invariant.
One way is to lift gauge parameters to physical fields on the boundary and treat them in a  chiral gauged WZW model.
The other  is to introduce $\mathcal{N}=(0,2)$ theories on the boundary with chiral fermions. \
 They induce gauge anomalies which compensate  non-invariant terms coming  from the three dimensional bulk theory under the gauge transformation

Next we evaluate the 3d-2d coupled indices on $S^1 \times \D$ in terms of the supersymmetric localization and try to relate the 3d-2d indices 
to holomorphic blocks. In the cases of Abelian gauge theories, we find that the 3d-2d indices reproduce holomorphic blocks with a  choice of the fugacity.

This paper is organized as follows. 
In section \ref{section2}, we study the super Yang-Mills action and the kinetic action of the chiral multiplet and 
introduce consistent boundary conditions for the $\mN=2$ multiplets. 
In section \ref{sec:BPSinteraction}, we discuss BPS boundary interactions for the superpotential term (three dimensional analogue of the matrix factorization) and  the Chern-Simons term. We also show the restriction of the $\mN=2$ supersymmetric transformation to the boundary torus leads to an $\mathcal{N}=(0,2)$ supersymmetric theory in two dimensions. 
In sections \ref{sec:localization} and \ref{Holomorphicblock}, we evaluate the one-loop determinants of 
three dimensional $\mathcal{N}=2$ (vector and chiral) multiplets and two dimensional  boundary
$\mathcal{N}=(0,2)$ (vector, chiral,  and Fermi) multiplets   
to study the relation between the 3d-2d indices on $S^1 \times D^2$ and holomorphic blocks in three dimensions.  
In section \ref{models}, we study properties of indices from  several  view points; first we consider  a three dimensional analogue of  the $\mathcal{N}=(2,2)$ hemisphere partition function of the $\C \mathbf{P}^N$-model on $D^2$ 
and point out the index of this model is related
 to the K-theoretic $J$-function of $\C \mathbf{P}^N$
and the $q$-deformed Whittaker function. 
Second we   study  the structure of the index of  $U(N)$ SQCD and 
its connection to K-theoretic vortex partition functions and surface operators.  
Third we study  the gauge/Bethe correspondence. We also analyze a domain wall 
on $S^1 \times S^2$ and evaluate  the 3d index with a domain wall.  
The last topic is the action of Wilson loops and vortex loops  on the 3d index.
The last section is devoted to summary and discussion. 

\section{$\mathcal{N}=2$ supersymmetric theory on $S^1 \times \D $}
\label{section2}
In this section, we will construct supersymmetric gauge theories on $S^1 \times \D $ and 
introduce supersymmetric boundary conditions for the hemisphere. The construction of  the supersymmetry and the Lagrangian 
is parallel  to the $S^1 \times S^2$ case \cite{Imamura:2011su}. 

The hemisphere $\D$ with the radius $r$ is specified by the  coordinates $(\mth, \vphi)$ 
with  $0 \le \mth \le \frac{\pi}{2}, 0 \le \vphi \le 2\pi $ and the boundary of the hemisphere is defined by $\mth=\frac{\pi}{2}$.
The circle $S^1$ is also parameterized by the coordinate $\tau $ with  $0 \le \tau \le \beta r$  
and  $\beta r$ is the circumference of $S^1$.
We can write the metric of $S^1 \times \D$ :
\bel
ds^2 =  r^2 d \mth^2+ r^2 \sin^2 \mth d \vphi^2+d \tau^2.
\ee
In the following, we use $\mu$'s $(\mu =1,2,3)$ for superscripts and subscripts in the curved space 
with $1=\mth, 2=\vphi $ and $3=\tau$. On the other hand, we take the symbol
$\hat{a}$ $(a=1,2,3)$ for variables in the local Lorentz frame.

Now we  construct supersymmetry  in the curved space that is realized  by conformal Killing spinors $\ep, \bep$.
These spinors should satisfy the set of equations:
\bel
\nabla_{\mu} \ep=\frac{1}{2r} \gamma_{\mu} \gamma_{3} \ep, \quad \nabla_{\mu} \bep=-\frac{1}{2r} \gamma_{\mu} \gamma_{3} \bep.
\ee
The solutions of these equations are given by
\bel
\ep= e^{\frac{\tau}{2r}}e^{-\frac{i}{2}  \gamma^2 \mth  } e^{\frac{i}{2} \gamma^3 \vphi}  \ep^{(2)}_o, \quad 
\bep= e^{-\frac{\tau}{2r}}e^{\frac{i}{2}  \gamma^2 \mth  } e^{\frac{i}{2} \gamma^3 \vphi}  \bep^{(2)}_o,\label{Killing}
\ee
where we choose the constant spinors $\ep^{(2)}_o=\gamma_{\hat{3}} \ep^{(2)}_o =(\ep_0, 0)^T$ 
and $\bep^{(2)}_o=- \gamma_3 \bep^{(2)}_o =(0, \bep_o)^T$
so that the component of the Killing vector $\bep \gamma^{\mu} \ep $ along the $\mth$-direction vanishes.
 
With the set of Killing spinors (\ref{Killing}), the 
supersymmetric transformation of the vector multiplet 
\eqref{SUSYtransvec1} is expressed as 
\bel
&&\delta A_{\mu}=\frac{i}{2} (\bar{\ep} \gamma_{\mu} \lambda - \bar{\lambda} \gamma_{\mu} \ep), \label{SUSYtransvec2}\\
&&\delta \sigma=\frac{1}{2} (\bar{\ep}  \lambda - \bar{\lambda}  \ep), \\
&&\delta \lambda=-\frac{1}{2} \gamma^{\mu \nu} F_{\mu \nu} \ep -D \ep+i \gamma^{\mu}D_{\mu} \sigma \ep
+\frac{i}{r} \sigma \gamma_{3}  \ep, \\
&&\delta \bar{\lambda}=-\frac{1}{2} \gamma^{\mu \nu} F_{\mu \nu} \bep +D \bep-i \gamma^{\mu}D_{\mu} \sigma \bep
+\frac{i}{r} \sigma \gamma_3 \bep, \\
&&\delta D=-\frac{i}{2} \bep \gamma^{\mu} D_{\mu} \lambda-\frac{i}{2} D_{\mu} \bar{\lambda} \gamma^{\mu}  \ep
+\frac{i}{2}[\bep \lambda, \sigma]+\frac{i}{2}[\bar{\lambda} \ep , \sigma]
+\frac{i}{4r} ( \bep \gamma_3 \lambda - \bar{\lambda} \gamma_{3}  \ep  ).
\label{SUSYtransvec3}
\ee
The Lagrangian density $\mathcal{L}_{\text{vec}}$ of super Yang-Mills theory is written in the $Q$-exact form:
\bel
2 \mathcal{L}_{\text{vec}}&=&\frac{1}{\ep_2 \ep_1} \delta_{\ep_2} \delta_{\ep_1} 
\mathrm{Tr}\Bigl[ \frac{1}{2} \lam \lam \Bigr] \non
&=&\mathrm{Tr} \Bigl[ \frac{1}{2}  F^{\mu \nu} F_{\mu \nu}+  D^{\mu} \sigma D_{\mu} \sigma+  D^2
+\frac{1}{r^2} \sigma^2 
-  \ep^{\mu \nu \rho}  F_{\mu \nu} D_{\rho} \sig
-\frac{1}{r}  \ep^{\mu \nu 3}  F_{\mu \nu} \sig 
+\frac{2}{r}  \sig  D_{3} \sig \Bigr] \non
&&+\mathrm{Tr}\Bigl[ i \lam \gamma^{\mu}  D_{\mu} \blam+i \lam [\blam,\sig]  -\frac{i}{2r} \blam  \gamma_{3 }    \lam \Bigr].
\label{SYMaction}
\ee
Here $\delta_{\epsilon_i} $ with $i=1,2$  denotes the supersymmetric transformation with $\bar{\epsilon}=0$ and $\epsilon=\epsilon_i$. 
We imposed  the following boundary condition for this vector multiplet at $\mth=\frac{\pi}{2}$:
\bel
&&\sig=0, \quad A_1=0, \quad \partial_1 A_{2}=0, \quad \partial_1 A_3=0, \quad D_1 (D -iD_{\hat{1}} \sig)=0, \non
&& \lam_1-\lam_2=0, \quad \blam_1-\blam_2=0, \quad \partial_{1} (\lam_1+\lam_2)=0, \quad  \partial_{1} (\blam_1+\blam_2)=0,
\label{veccondition} 
\ee
with $\lam=(\lam_1, \lam_2)^T$ and $\blam=(\blam_1, \blam_2)^T$. 
This boundary condition is compatible with the supersymmetric transformation 
 (\ref{SUSYtransvec2})-(\ref{SUSYtransvec3}).
When we impose the above boundary condition, induced surface terms vanish and   the supercurrents do not have any components in the normal  directions to 
the boundary.

Next we consider the chiral multiplet $(\phi,\psi, F)$. The supersymmetric transformation of the chiral multiplet is given by
\bel
&&\delta \phi= \bep \psi ,\label{SUSYchi2} \\
&&\delta \bphi= \ep \bpsi ,\\
&&\delta \psi= i \gamma^{\mu} \ep D_{\mu} \phi+i \ep \sig \phi+\frac{i\Delta}{r} \gamma_{3}  \ep \phi+\bep F ,\\
&&\delta \bpsi= i \gamma^{\mu} \bep D_{\mu} \bphi+i \bphi \sig \bep-\frac{i\Delta}{r} \bphi \gamma_{3}  \bep +\bF \ep  ,\\
&&\delta F=\ep (i \gamma^{\mu} D_{\mu} \psi-i \sig \psi -i\lam \phi)+\frac{i}{2r} (2\Delta-1) \ep \gamma_{3} \psi ,\\
&&\delta \bF=\bep (i \gamma^{\mu} D_{\mu} \bpsi-i \bpsi \sig  +i \bphi \bar{\lam} )-\frac{i}{2r} (2\Delta-1) \bep \gamma_{3} \bpsi .
\label{SUSYchi3}
\ee
Here $\Delta$ is the R-charge of $\phi$.
The kinetic term of the chiral multiplet is also given in the $Q$-exact form:
\bel
 \mathcal{L}_{\text{chi}}&=&\frac{1}{\ep_2 \ep_1} \delta_{\ep_2} \delta_{\ep_1} \left( \bphi F \right) \non
&=&-\bphi D^{\mu} D_{\mu}   \phi+\bphi \sig^2 \phi+i \bphi D \phi+\bF F +\frac{1-2\Delta}{r}  \bphi D_3 \phi +\frac{\Delta-\Delta^2}{r^2} \bphi \phi  \non
&&- i  \bpsi \gamma^{\mu} D_{\mu}  \psi  +i \bpsi \sig \psi    
+\frac{i(1-2\Delta)}{2r}  \bpsi  \gamma_{3}  \psi
+i \bpsi  \lam \phi-i \bphi \bar{\lam}  \psi. 
\label{Lagchiral}  
\ee
We can introduce the 
Neumann boundary condition for the chiral multiplet at $\mth=\frac{\pi}{2}$:
\bel
&&\partial_{1} \phi=0, \quad \partial_{1} \bphi=0, \quad F=0, \quad \bar{F}=0 ,\non
&&\psi_1+\psi_2=0, \quad \bpsi_1+\bpsi_2=0, \quad \partial_{1} (\psi_1-\psi_2 )=0, \quad \partial_{1} (\bpsi_1-\bpsi_2)=0,
\label{Ncondition}
\ee 
with $\psi=(\psi_1, \psi_2)^T$ and $\bpsi=(\bpsi_1, \bpsi_2)^T$.  

The Dirichlet boundary condition is given by
\bel
&&\phi=0, \quad  \bphi=0, \quad \partial_{1} (i e^{\frac{\tau}{r}} e^{i \varphi} \partial_{\hat{1}} \phi +F)=0, 
\quad \partial_{1} (i e^{-\frac{\tau}{r}} e^{-i \varphi} \partial_{\hat{1}} \bphi +\bar{F})=0  ,\non
&&\psi_1-\psi_2=0, \quad  \bpsi_1-\bpsi_2=0, \quad  \partial_1(\psi_1+\psi_2)=0, \quad  \partial_1(\psi_1+\psi_2)=0.  
\label{Dcondition}
\ee 
The Neumann (Dirichlet) boundary condition (\ref{Ncondition}), (\ref{Dcondition})  
is the 3d analogue of the Neumann (Dirichlet)  boundary condition in two dimensional $\mathcal{N}=(2,2)$
theories on $\D$ \cite{Honda:2013uca, Hori:2013ika} and different from the boundary condition imposed in \cite{Sugishita:2013jca}.

\section{The  BPS boundary interactions on the torus}
\label{sec:BPSinteraction}
Under the boundary conditions (\ref{veccondition}) and  (\ref{Ncondition}),  the Lagrangians (\ref{SYMaction}) and  (\ref{Lagchiral}) are invariant  under the supersymmetric transformations generated by $\delta_{\ep}, \delta_{\bep}$. 
On the other hand, the supersymmetric transformations of the supersymmetric Chern-Simons term and 
the superpotential term do not vanish and 
we have to introduce supersymmetric boundary interactions  
 to cancel surface terms coming from these terms.  

\subsection{Three dimensional analogue of matrix factorization}
In the presence of the boundary, the supersymmetric variation of the superpotential does not vanish.
In two dimensional $\mathcal{N}=(2,2)$ theories, the boundary term for the superpotential is cancelled by 
the boundary interaction which satisfies the matrix factorization \cite{Kapustin:2002bi, Herbst:2008jq}. 
The three dimensional analogue of the matrix factorization was  pointed out first in \cite{Gadde:2013sca}.
In this subsection, we study the matrix factorization for the three dimensional $\mN=2$ theories on $S^1 \times \D$.

The supersymmetric transformations of the superpotentials $W$ and $\bar{W}$  induce the following surface terms on the boundary:
\bel
\delta S_{W}=\int_{T^2} \sqrt{g} d^2x  \sum_{I} \left( \ep \gamma_{\hat{1}} \psi_I   \frac{\partial W}{\partial \phi_I}+ \bep \gamma_{\hat{1}} {\bpsi}_I \frac{ \bar{W}}{\partial \bar{\phi}_I}  \right).
\label{superpot}
\ee
Here $T^2$ is the boundary of $S^1 \times D^2$ and   $I$ in the sum labels the chiral multiplets with the Neumann boundary condition in the theory.
We have to introduce boundary interactions which compensate the above boundary terms.
Now we restrict the supersymmetric transformations of the three dimensional $\mathcal{N}=2$ multiplets to the boundary ($\mth=\frac{\pi}{2}$) 
in order to construct the BPS and gauge invariant boundary interactions.
First of all, we will consider the vector multiplet 
(\ref{SUSYtransvec2})-(\ref{SUSYtransvec3}) restricted on the boundary.
Under the condition  (\ref{veccondition}), associated supersymmetric transformations  are given by
\bel
\delta A_{\hat{2}}&&=\frac{i}{2} (\bar{\ep} \gamma_{\hat{2}} \lambda - \bar{\lambda} \gamma_{\hat{2}} \ep) 
= \bar{\ep}'  \lambda_1 - \bar{\lambda}_1  \ep',  \non
\delta A_{\hat{3}}&&=\frac{i}{2} (\bar{\ep} \gamma_{\hat{3}} \lambda - \bar{\lambda} \gamma_{\hat{3}} \ep) 
= i( \bar{\ep}'  \lambda_1 - \bar{\lambda}_1  \ep' ), \non
\delta \lambda&&=- i F_{\hat{2} \hat{3}} \ep -\hat{D} \ep, \label{02vectersusy}\\
\delta \bar{\lambda}&&=- i F_{\hat{2} \hat{3}} \bep +\hat{D} \bep, \non
\delta(-\hat{D}-i F_{\hat{2} \hat{3}})&&=2 \bep' ( D_{\hat{2}}+i D_{\hat{3}})  \lam_1  -\frac{2i}{r} \bep'  \lam_1, \non
\delta(-\hat{D}+i F_{\hat{2} \hat{3}})&&=-2 \ep' (D_{\hat{2}} +i D_{\hat{3}}) \blam_1   -\frac{2i}{r} \ep' \blam_1 \nonumber.  
\ee
Here we defined  $\hat{D}:=D-i D_{\hat{1}} \sigma$ and $\ep':=\frac{e^{\frac{\tau}{2r}} e^{i\frac{\vphi}{2}}}{\sqrt{2}} \ep_o, \bep':=\frac{e^{-\frac{\tau}{2r}} e^{-i\frac{\vphi}{2}}}{\sqrt{2}} \bep_o$. 
The commutation relations of these transformations are summarized in  appendix B.
In the flat space limit $r \to \infty$, the set of the above transformations
becomes that of  two dimensional $\mathcal{N}=(0,2)$  vector multiplet. We shall call
this multiplet a boundary $\mathcal{N}=(0,2)$  vector multiplet.
A  Lagrangian  can be expressed as a $\delta_{\epsilon'}$-exact form and 
invariant under the supersymmetric transformations (\ref{02vectersusy}):
\bel
\ep' \mathcal{L}^{\mathcal{N}=(0,2)}_{\text{vec}}&&=\delta_{\ep'} \mathrm{Tr} (-\hat{D}+iF_{\hat{2} \hat{3}}) \lam_1  \non
&&=\ep' \left(F^2_{\hat{2} \hat{3}}+\hat{D}^2+2 \blam_1 (D_{\hat{2}}+i D_{\hat{3}}) \lam_1-\frac{2i}{r} \blam_1 \lam_1\right).
\ee 

When we take the Neumann boundary condition, the supersymmetric transformation of the chiral multiplet on the boundary is given by
\bel
&&\delta \phi= \bep \psi = \bep' \psi', \non
&&\delta \bphi= \ep \bpsi = \ep' \bpsi',\non
&&\delta \psi= \gamma_{\hat{3}} \ep ( D_{\hat{2}}+iD_{\hat{3}} ) \phi+\frac{i\Delta}{r} \gamma_{3}  \ep \phi, \\
&&\delta \bpsi= \gamma_{\hat{3}}  \bep ( D_{\hat{2}}+iD_{\hat{3}} )  \bphi-\frac{i\Delta}{r} \bphi \gamma_{3}  \bep, \nonumber
\label{02chiralsusy}
\ee   
where we defined $\psi':=\psi_1-\psi_2, \bpsi':=\bpsi_1-\bpsi_2$.
When we take the limit $r \to \infty$, the above transformation  becomes the 
$\mathcal{N}=(0,2)$ supersymmetric transformation of the boundary chiral multiplet $(\phi, \psi')$.
Then we introduce the Lagrangian  as a $\delta_{\epsilon'}$-exact form:
\bel
\ep' \mathcal{L}^{\mathcal{N}=(0,2)}_{\text{chi}}&&=\delta_{\ep'} \left(\frac{1}{2}\bphi (D_{\hat{2}}-iD_{\hat{3}}) \psi^{'} + i\bphi \lam_1 \phi \right) \non
&&= \ep' \Bigl( \bphi (D_{\hat{2}}-iD_{\hat{3}}) (D_{\hat{2}}+iD_{\hat{3}}) \phi+\frac{1}{2} \bpsi ' (D_{\hat{2}}-iD_{\hat{3}})  \psi '  
\non
&& \qquad \qquad +\frac{ i\Delta }{r} \bphi (D_{\hat{2}}-iD_{\hat{3}})  \phi   +  
i\bphi \blam_1 \psi ' + i \bpsi_1 \lam_1 \phi + \bphi (F_{\hat{2} \hat{3}} -i\hat{D}) \phi \Bigr) . 
\ee
If we take $r \to \infty$, $\mathcal{L}^{\mathcal{N}=(0,2)}_{\text{chi}}$ is reduced to the Lagrangian for the $\mathcal{N}=(0,2)$ chiral multiplet in the flat space.
These $\mathcal{N}=(0,2)$ boundary (vector and chiral) multiplets are constructed  through  restriction of the bulk supersymmetry on the boundary. 
In addition, there are  new multiplets in the $\mathcal{N}=(0,2)$ theory characterized by holomorphic functions $E(\phi)$'s,  namely  Fermi multiplets and we can  
construct the boundary interactions which cancel the variation of the superpotential term.  
We shall introduce the Fermi multiplet $(\Psi, G)$ on $T^2$ 
coupled to the boundary  $\mathcal{N}=(0,2)$ (vector and chiral) multiplets in
the supersymmetric way.  
The supersymmetric transformation of the boundary Fermi multiplet is given by
\bel
&&\delta \Psi= 2E \ep'+2 \bep' G ,\label{SUSYFermi1} \\
&&\delta \bar{\Psi}=2 \bar{E} \bep'  + 2 \ep' \bar{G}  ,\\
&&\delta G=- \ep' \psi_{E}+ \ep' ( D_{\hat{2}} +i D_{\hat{3}} )\Psi + \frac{i}{r} (\tilde{\Delta}-1) \ep'  \Psi ,\\
&&\delta \bar{G}=- \bep' \bpsi_{E}+ \bep' ( D_{\hat{2}} +i D_{\hat{3}} ) \bar{\Psi} + \frac{i}{r} (1-\tilde{\Delta}) \bep' \bar{\Psi}.
\label{SUSYFermi2}
\ee
Here $\psi_{E}:=\sum_{I} \frac{\partial E(\phi)}{\partial \phi_I} {\psi'}_{I}, \bpsi_{E}:=\sum_{I} \frac{\partial {E}(\bphi)}{\partial \bphi_I} {\bpsi'}_{I}$ 
and $(\phi_I, {\psi'}_I)$'s are $\mathcal{N}=(0,2)$ boundary chiral multiplets. 
 $\sum_I$ runs over the  $\mathcal{N}=(0,2)$ chiral multiplets   in $E(\phi)$ and 
we also require  the relation $\sum_{I  } \Delta_I =\tilde{\Delta} $. 
One can show  that commutators of these supersymmetries  generate a translation, R-symmetry and gauge transformations.

Next we study the Dirichlet boundary condition.
We  remark that 
the supersymmetric transformations (\ref{SUSYchi2})-(\ref{SUSYchi3}) with the Dirichlet boundary condition on the boundary torus lead to 
(\ref{SUSYFermi1})-(\ref{SUSYFermi2}) with $E(\phi)=0$ by the following  redefinition:
\bel
&&G:=i e^{ \frac{\tau}{r}} e^{i \varphi} D_{\hat{1}} \phi  +F, \quad \bar{G}:=i e^{\frac{-\tau}{r}} e^{-i \varphi} D_{\hat{1}}  \bphi +\bar{F}, \non
&&\Psi:=\psi_1, \quad \bar{\Psi}:=\bpsi_1, \quad \tilde{\Delta}=2\Delta.
\ee
This is the special case of the $\mathcal{N}=(0,2)$ boundary Fermi multiplet.

Now we return to the boundary Fermi multiplet. 
The Lagrangian of the boundary Fermi multiplet can be constructed as
\bel
 \ep' \mathcal{L}^{\mathcal{N}=(0,2)}_{\text{Fermi}}&&=\delta_{\ep'} (\bar{\Psi} G+\bar{E} \Psi ) \non
&&= \ep' \left(    -  \bar{\Psi} ( D_{\hat{2}}+i D_{\hat{3}} ) \Psi + 2\bar{G} G + 2\bar{E} E -  \bpsi_{E} \Psi-\bar{\Psi} \psi_{E}
 + \frac{i}{r} (  1 - \tilde{\Delta} ) \bar{\Psi} \Psi
 \right). \non
\ee  
We can also introduce the  potential term for this multiplet
\bel
\mathcal{L}_{J}=\sum_a \left( G_a J^a-\frac{1}{2} \Psi_{a} \psi_{J^a} \right)+(\text{c.c}),
\ee
which induces the following terms on  $T^2$ through the variation
\bel
\int_{T^2} \sqrt{g} d^2x  \delta \mathcal{L}_{J} =
\int_{T^2} \sqrt{g} d^2x  \sum_{I,a } \left( - \ep ' {\psi'}_{I} \frac{\partial (E_a J^a) }{ \partial \phi_I} +(\text{c.c}) \right)  .
\label{fsuperpot}
\ee 
Here $(\text{c.c})$ denotes the complex conjugation. 
By comparing this with the term (\ref{superpot}), we find that 
cancellation between (\ref{superpot}) and (\ref{fsuperpot}) occurs when 
the relation $\sum_a E_a J^a=W$ is satisfied.
We further require each monomial in $ E_a J^a$ has R-charge $2$ so that 
the relation $\sum_a E_a J^a=W$ can be regarded as {\em the three dimensional matrix factorization}.

\subsection{Supersymmetric Chern-Simons term}
In this subsection, we consider  $\mathcal{N}=2$ supersymmetric  Chern-Simons theory and investigate 
boundary terms induced by variations of the bulk action.
The boundary effect is first studied   for   $\mathcal{N}=1$ Chern-Simons theory \cite{Sakai:1989nh}
and  $\mathcal{N}=2$ Abelian case is also studied in  \cite{Berman:2009kj}.

We first treat the Chern-Simons term  with the gauge group $G$ and construct consistent boundary interactions. 
Later we treat gauge-flavor mixed Chern-Simons terms.
The $\mathcal{N}=2$ Chern-Simons theory has a gauge field $A_{\mu}$,  bosonic fields $D$,$\sigma$ and fermions $\lambda$, $\bar{\lambda}$. 
These fields take  values in the Lie algebra of the gauge group  $G$ and transform in the adjoint representation. 
 This model has a 
 coupling constant $\kappa$ called a Chern-Simons level  and Chern-Simons action is 
 written by
\bel
&&S_{CS}=\frac{i\kappa}{4\pi } \int d^3 x  \,\mathcal{L}_{CS},\non
&&\mathcal{L}_{CS} =
 \varepsilon^{\mu \nu \rho}  \mathrm{Tr}  
\left(\partial_{\mu} A_{\nu} A_{\rho} +\frac{2i}{3} A_{\mu} A_{\nu} A_{\rho} \right) 
+\sqrt{g}\,\mathrm{Tr}(-\blam \lam +2 \sig D), 
\ee
where $\varepsilon^{\mu\nu\rho}$ is an antisymmetric tensor density.
The supersymmetric variation of this term is evaluated as 
\bel
\delta \mathcal{L}_{CS}=\frac{i}{2} \partial_{\mu}
 \mathrm{Tr}  \varepsilon^{\mu \nu \rho} \Bigl(   \bep \gamma_{\nu} \lam  A_{\rho} -\bar{\lambda} \gamma_{\nu} \ep  A_{\rho}  \Bigr)
-i \partial_{\mu}\mathrm{Tr} \sqrt{g} \Bigl(  \bep \gamma^{\mu} \lam  \sig +\bar{\lambda} \gamma^{\mu}  \ep \sig  \Bigr).
\ee 
On $S^1 \times \D$ with the boundary condition (\ref{veccondition}), the second term in the above equation 
can be dropped out and the first term 
 leads to the following boundary term
\bel
&&S_{CS}=
-\frac{i \kappa}{8\pi} \int_{T^2} 
\sqrt{g^{ (2)}}
 d^2x \sum_{\mu=2,3} \mbox{Tr}[A_{\mu}
( \bep \gamma^{\mu} \lam  -\bar{\lambda} \gamma^{\mu} \ep  ) ],
\ee
where $\sqrt{g^{(2)}}$ is the measure of the two dimensional boundary and 
is related to the bulk three dimensional one by $\sqrt{g}=r\sqrt{g^{(2)}}$.
By introducing a boundary Chern-Simons term 
\bel
&&S_{b.CS}=\frac{- \kappa}{8\pi}  \sum_{\mu=2,3} \int_{T^2} 
\sqrt{g^{(2)}} d^2x \,  \mathrm{Tr} (  A_{\mu}  A^{\mu} ),
\label{SUSYCSb} 
\ee 
we find that $S_{CS}+S_{b.CS}$ preserves the supersymmetry generated by $\delta_{\ep'}, \delta_{\bep'}$.
But $S_{CS}+S_{b.CS}$  still breaks gauge invariance in the presence of the boundary and 
we have to resolve this problem. 
There are two choices to recover the gauge symmetry: 
\begin{itemize}
\item[(i)] One introduces some boundary $\mathcal{N}=(0,2)$ multiplets to induce anomaly inflows. 

\item[(ii)] One treats the gauge degrees of freedom as physical fields  on the boundary and couples them with a chiral gauged WZW  model.
\end{itemize}
We first explain the choice (i).  The gauge non-invariant term of  $S_{CS}+S_{b.CS}$ is compensated by gauge anomalies from the boundary
 $\mN=(0,2)$ chiral and Fermi multiplets when the following condition is satisfied: 
 \bel
\kappa \mathrm{Tr} (T^a T^b) =\sum_{m:\text{chiral}}\mathrm{Tr}_{\mathcal{R}_m} (T^a T^b) -
\sum_{n:\text{Fermi}}\mathrm{Tr}_{{\mathcal{R}}_n} (T^a T^b) .
\label{treeanomaly}
\ee
The left hand side comes from  an infinitesimal gauge transformation of $S_{CS}+S_{b.CS}$ and the symbol ``$\mathrm{Tr}$'' denotes the trace over the Lie algebra  of $G$. 
 In the right hand side,  $\mathrm{Tr}_{\mathcal{R}_m} (T^a T^b)$, (resp. $\mathrm{Tr}_ {{\mathcal{R} }_n} (T^a T^b)$ )
 comes  from the gauge anomaly coefficient of the chiral (resp. Fermi) multiplet in the representation $\mathcal{R}_m$, (resp. ${\mathcal{R}}_n$). 
The sums $\sum_{m:\text{chiral}}$ and $\sum_{n:\text{Fermi}}$ run over the number of boundary chiral and Fermi multiplets, respectively.
It means the classical gauge anomaly (the left hand side) and  the one-loop gauge anomalies (the right hand side) cancel each other.
This type  of cancellation was  also considered in the context of an $\mathcal{N}=(0,2)$ gauged WZW model \cite{Berglund:1995dv}.

But there is a subtle point; in the case of $\mathcal{N}=2$ theories in three dimensions, the bare Chern-Simons level is shifted by 
one-loop effects of 3d dynamical fermions. This implies that the coefficient of the boundary term, i.e.,  the mass term of the 2d gauge field is also shifted. 
In two dimensions, it is  known that the gauge field acquires a mass term by the quantum effect \cite{Schwinger:1962tp} and  the coefficient of the  boundary Chern-Simons term is also shifted
 by the one-loop effects of 2d fermions.
 Then the  
left hand side of (\ref{treeanomaly}) is expected to be replaced by some effective Chern-Simons level. 
We will observe how this level shift appears in the context of  the localization computation in section \ref{Holomorphicblock}. 

Next we shall consider quiver Chern-Simons theories, where gauge fields are not necessarily dynamical. 
In fact, there are possible mixing terms between  the dynamical gauge fields and the background flavor gauge fields, 
namely, mixed  Chern-Simons terms. 
For simplicity, we assume the group is Abelian, but it is straightforward to generalize to non-Abelian cases.
We take  the gauge and flavor groups as $U(1)^{N}_G \times U(1)^{N_f}_F $; the dynamical gauge group $U(1)^{N}_G $,
and the flavor symmetry group  $U(1)^{N_f}_F$. 
Further there are other possible mixings, for example, 
the dynamical gauge symmetry and the $U(1)_R$ R-symmetry,  or the flavor symmetry and the $U(1)_R$ R-symmetry. 
We will discuss  such  mixed Chern-Simons terms in section 5.

  $\mathcal{N}=2$  Abelian quiver theory  is described by the action
\bel
S_{CS}=\sum_{s,t=1}^{N+N_f} \frac{i \kappa_{s t}}{4\pi }\int_{S^1 \times \D}  d^3 x \left[ \varepsilon^{\mu \nu \rho} \partial_{\mu} A^{(s)}_{\nu} A^{(t)}_{\rho} +
\sqrt{g}( -\blam^{(s)} \lam^{(t)} +2 \sig^{(s)} D^{(t)})   \right].
\ee
$\kappa_{s t}$ with $s, t=1,\cdots, N+N_f$ are the gauge-flavor mixed Chern-Simons levels.
We will collect  these levels into 
one symmetric $(N+N_f) \times (N+N_f)$ matrix  $\kappa_{s t}$. 
$A^{(s)}_{\mu}$ represents the collection of $U(1)^N_G$ dynamical gauge fields and $U(1)^{N_f}_F $ background gauge fields.
As in the case of the non-Abelian theory (\ref{SUSYCSb}), the supersymmetric  boundary   term is given by
\bel
S_{b.CS}&&=\sum_{s,t=1}^{N+N_f} \frac{-\kappa_{s t}}{2 \pi }\int_{T^2} \sqrt{g} d^2x    A^{(s)}_{z}  A^{(t)}_{\bar{z}} .
\label{quivermixedbt}
\ee
Here we defined $A_{z}:=(A_{2}-i A_{3})/2$ and $A_{\bar{z}}:=(A_{2}+i A_{3})/2$. 
The infinitesimal $U(1)^{N}_G \times U(1)_{F}^{N_f}$  transformation of the mixed Chern-Simons term
 leads to non-invariant terms on the boundary:
 \bel
 \delta_{\alpha} (S_{CS} +S_{b.CS} )=\sum_{s,t=1}^{N+N_f} \frac{\kappa_{s t}}{2\pi}\int_{T^2} F^{(s)}_{z \bar{z}} \alpha^{(t)} .
\label{3dmixedanomaly} 
\ee
The gauge-gauge and gauge-flavor mixed parts in (\ref{3dmixedanomaly})  can be  cancelled from
 corresponding  mixed anomalies in the boundary $\mN=(0,2)$ (chiral and Fermi) multiplets
 when the following conditions are satisfied: 
\bel
\kappa_{st}=\sum_{m: \text{2d chiral}}Q^{m}_{\text{2d},s} Q^{m}_{\text{2d},t} - \sum_{n:\text{2d.Fermi}}\tilde{Q}^{n}_{\text{2d},s} \tilde{Q}^{n}_{\text{2d},t} .
\label{mixedinflow}
\ee
Here $Q^m_{\text{2d}, s}$ denotes the charge of  the $s$-th $U(1)$ in $U(1)^{N}_G \times U(1)^{N_f}_F$ for the $m$-th $\mN=(0,2)$ chiral multiplet.  
$\tilde{Q}^n_{\text{2d},s}$ denotes   the charge of  the $s$-th $U(1)$  in $U(1)^{N}_G \times U(1)^{N_f}_F$ for  the $n$-th $\mN=(0,2)$ Fermi multiplet.  
The sum of each term  runs over the $\mN=(0,2) $ chiral and Fermi multiplets respectively.
If the bare Chern-Simons levels $\kappa$'s are replaced by the effective Chern-Simons levels, 
 the equation (\ref{mixedinflow}) matches with the condition of 
Chern-Simons levels  in the context of the holomorphic blocks \cite{Beem:2012mb}(see also \cite{Gadde:2013sca}).

Next  we explain the second choice (ii).
Under the finite  gauge transformation $A^{g}_{\mu} = i \partial_{\mu} g g^{-1} +g A_{\mu} g^{-1}, ( g \in G$), 
the Chern-Simons term  and the supersymmetric boundary term transform as
\bel
\varepsilon^{\mu \nu \rho} \mathrm{Tr}\left(\partial_{\mu} A_{\nu} A_{\rho} +\frac{2i}{3} A_{\mu} A_{\nu} A_{\rho} \right)^g
&&=  \varepsilon^{\mu \nu \rho} \mathrm{Tr} \Bigl( -\frac{1}{3} \partial_{\mu} g g^{-1} \partial_{\nu} g g^{-1} \partial_{\rho} g g^{-1}  
+i  \partial_{\mu} ( A_{\nu} g^{-1} \partial_{\rho} g )  \Bigr) \non
&& \quad \qquad +\varepsilon^{\mu \nu \rho} \mathrm{Tr} \Bigl(\partial_{\mu} A_{\nu} A_{\rho} +\frac{2i}{3} A_{\mu} A_{\nu} A_{\rho} \Bigr) ,
\ee
\bel
\mathrm{Tr}(A_{z}  A_{\bar{z}})^g
&&=\mathrm{Tr} \Bigl(-\partial_{z} g g^{-1} \partial_{\bar{z}} g g^{-1}+i g^{-1} \partial_{z} g A_{\bar{z}} +i A_{z} g^{-1} \partial_{\bar{z}} g +A_{z}  A_{\bar{z}} \Bigr).
\ee
Then the transformation of the combined  action is given by
\bel
(S_{CS}+S_{b.CS})^g =S_{CS}+S_{b.CS} -S_{c.GWZW},
\ee
with
\bel
S_{c.GWZW}[g, A_{\bar{z}}] &&=-\frac{\kappa}{2\pi } \int_{T^2} \mathrm{Tr}  \bigl(\partial_{z} g g^{-1} \partial_{\bar{z}} g g^{-1} \bigr) 
+\frac{i \kappa}{12\pi} \int_{S^1 \times \D}  \varepsilon^{\mu \nu \rho} 
\mathrm{Tr} \bigl( \partial_{\mu} g g^{-1} \partial_{\nu} g g^{-1} \partial_{\rho} g g^{-1} \bigr) 
\non && \quad +\frac{i \kappa}{\pi}  \int_{T^2} \mathrm{Tr} \bigl(  g^{-1} \partial_{z} g A_{\bar{z}} \bigr).
\label{GWZW}
\ee
The first line in  (\ref{GWZW}) is the action of $G$ Wess-Zumino-Witten (WZW) model and the second line is the chiral $G/G$ gauged term with a 
right  $G$-action. 
The supersymmetric transformation trivially  acts on the $G$-elements and (\ref{GWZW})  is invariant by the supersymmetric transformation,  
since the fermionic superpartner of the chiral $G/G'$ gauged WZW model takes  value in the subspace orthogonal 
to $\mathrm{Lie}(G')$ in $\mathrm{Lie}(G)$.
It is this combination of gauge fields $A_{\bar{z}}$
that is invariant under supersymmetric transformations.

The combination $ S_{CS}+S_{b.CS} +S_{c.GWZW} $ is also gauge invariant 
because  we have identities under  the right $g'$-action  with $g \to g g', (g, g' \in G) $
\bel
&&(S_{CS}+S_{b.CS})^{g'}=S_{CS}+S_{b.CS} -S_{c.GWZW}[g', A_{\bar{z}}], \\
&&S_{c.GWZW}[g g', A^{g'}_{\bar{z}}] =S_{c.GWZW}[g, A_{\bar{z}}] +S_{c.GWZW}[g', A_{\bar{z}}].
\ee
This is called the chiral gauged WZW model. 
When we treat the gauge degrees of freedom as physical fields,
 we find that $S_{CS}+S_{b.CS} +S_{c.GWZW}$ is fully invariant under the $G$ gauge transformation.
The partition function of this WZW model with the background gauge field  
is known as  the holomorphic wave functional $\Psi_{CS}[A_{\bar{z}}]$  in the pure Chern-Simons theory \cite{Elitzur:1989nr}. 

It is possible to consider more general  boundary interactions  by gauging subgroup $H$ of $G$ and considering 
a left $H$-action :
$g \to h^{-1} g g', (g, g' \in G, h \in H)$.  Although the chiral gauged WZW model  is anomalous under the left $H$- and right $G$-actions, 
we can  cancel the  gauge anomaly for $H$ by introducing appropriate $\mathcal{N}=(0,2)$ (chiral and Fermi) multiplets coupled to 
the $\text{Lie}(H)$-valued gauge field. Namely, when 
these   $\mathcal{N}=(0,2)$ multiplets satisfy some suitable relations to cancel the anomaly,  the theory becomes consistent. 
It is interesting to study these boundary interactions in detail and evaluate  such 3d-2d coupled partition functions. 
But in the rest of this article, we mainly consider the case (i). 

\section{Index on $S^1 \times \D$ and localization}
\label{sec:localization}
In this section, we evaluate the partition function (index) on $S^1 \times \D$ via localization. 
On $S^1 \times S^2$,  the supersymmetric variation parameters $\epsilon, \bar{\epsilon} $ cannot be periodic  along the $S^1$ direction  
\cite{Imamura:2011su}, then we  impose the twisted  boundary condition along this $S^1$ direction:
\bel
\Phi (\tau + \beta r) =e^{ (-{ \sf J}_3-{\sf R})\beta_1+{\sf J_3} \beta_2 +\sum_{l} F_l M_l} \Phi (\tau),  \quad (\beta=\beta_1+\beta_2).
\label{twistedPB} 
\ee
Here $\Phi$ is a field in the supermultiplets, ${\sf J_3}$ is the generator of a rotation along the $\vphi$-direction, ${\sf R}$ is the $U(1)$ R-charge and 
$F_l$'s are charges of global symmetry groups.
$\beta_1$, $\beta_2$ and $M_l$'s are fugacities for these charges.  
Under the twisted boundary condition, the partition function on $S^1 \times S^2$ defines a superconformal index:   
\bel
\mathcal{I}_{S^1 \times S^2}=\mathrm{tr}_{\mathcal{H}(S^2)} \left[ (-1)^F e^{-\beta_1({\sf D}-{\sf R}-{\sf J_3})} e^{-\beta_2({\sf D}+{\sf J_3})} 
e^{-\sum_{l} F_{l} M_{l}}\right].
\label{eq:SUSYind0}
\ee 
Here ${\sf D}$ is the generator of the translation along the $S^1$-direction. 
The superconformal index  counts the number of BPS operators which saturate
the bound ${\sf D}-{\sf R}-{\sf J_3} \ge 0$ and does not depend on the  fugacity $\beta_1$. 

In a similar way, the partition function on $S^1 \times D^2$ with the boundary conditions (\ref{veccondition}),(\ref{Ncondition}) or (\ref{Dcondition}) 
defines  the following  index: 
\bel
\mathcal{I}_{S^1 \times D^2}&=&\mathrm{tr}_{\mathcal{H}(D^2)}
 \left[ (-1)^F e^{-\beta_1({\sf D}-{\sf R}-{\sf J_3})} e^{-\beta_2({\sf D}+{\sf J_3})} e^{-\sum_l  F_{l} M_{l}}\right] \non
&=&\mathrm{tr}_{\mathcal{H}(D^2)}
 \left[ (-1)^F  e^{-\beta_2({\sf R}+{\sf 2 J_3})} e^{-\sum_l  F_{l} M_{l}}\right].
\label{eq:SUSYind}
\ee
We will show  that all the one-loop determinants on  $S^1 \times \D$ do not depend on $\beta_1$, which means 
that \eqref{eq:SUSYind} counts the number of  operators saturating the bound ${\sf D}-{\sf R}-{\sf J_3} \ge 0$ on $S^1 \times \D$.

Recently the indices on $S^1 \times S^2_b$ and $S^1 \times \R \mathbf{P}^2$ have been studied  in \cite{Tanaka:2014oda}.
These are trigonometric  deformations of  
the partition functions on the two dimensional squashed sphere $S^2_{b}$ \cite{Gomis:2012wy} and on
the real projective space $\R \mathbf{P}^2$ \cite{Kim:2013ola} by the Kaluza-Klein modes of $S^1$. 
The index on $S^1 \times S^2_b$ 
 does not depend on 
the squashing parameter $b$ and agrees  with the ordinary superconformal index.  
At least, the (three dimensional) bulk part of the index on $S^1 \times D^2$ 
is also expected to be independent of the squashing deformation of the two dimensional hemisphere.
Moreover  the index on $S^1 \times D^2$ does not explicitly depend on the radius $r$ 
of the two dimensional hemisphere,
since the radius of the hemisphere enters in the index on $S^1 \times \D$ only through the matrix integral variables 
when the localization method is applied. 
When the boundary interactions are absent, the index on $S^1 \times D^2$ is expected to be independent of  both the squashing parameter and the radius of the hemisphere.
Then the index on $S^1 \times D^2$ can be  identified with 
 an index on the circle times the two dimensional flat space,  $\mathcal{I}_{S^1 \times \D} \simeq \mathcal{I}_{S^1 \times \R^2}$:
\bel
\mathcal{I}_{S^1 \times D^2}&=&\mathrm{tr}_{\mathcal{H}(\R^2)}
 \left[ (-1)^F  e^{-\beta_2({\sf R}+{\sf 2 J_3})} e^{-\sum_{l} F_{l} M_{l}}\right].
\ee
This is a BPS index of three dimensional $\mathcal{N}=2$ supersymmetric theory. 
Our 3d-2d index should be  thought of as a three dimensional BPS index with boundary interactions.

In the calculation of the localization 
of  super Yang-Mills Lagrangian (\ref{SYMaction}), 
we find that the zero locus (saddle point) of  the $Q$-exact action is given by 
\bel
A_3=\text{constant}=a=\sum_{c=1}^{N} a^c H^c, 
\label{saddlepoint}
\ee
and the other fields are trivial. $\{ H^c\}_{c=1}^{N }$ denotes the set of generators of the Cartan subalgebra of $\text{Lie}(G)$ and $N$ is the rank of $\text{Lie}(G)$.
The boundary Chern-Simons term 
 at the saddle point \eqref{saddlepoint}  is given by 
\bel
\exp \left(S_{b.CS} \right)=\exp \left( - \frac{\kappa}{4\beta} \mathrm{Tr} (i \beta r a)^2 \right).
\label{ClassicalCS}
\ee
 Note that (\ref{ClassicalCS})  is not invariant under  the large gauge transformation $i \beta r a \to i \beta r a  +\sum_{c=1}^N d^c H^c$ with $d_c \in \mathbb{Z}$.  

As another possibility of a $Q$-closed form, we have the FI-term
\bel
\mathcal{L}_{\text{FI}}= i \zeta \mathrm{Tr} \left( \frac{A_3}{r} - D\right).
\label{FIterm}
\ee
In fact, the supersymmetric transformation of this FI-term does not have surface terms and becomes 
a $Q$-closed form
 under the boundary condition (\ref{veccondition}).
This FI-term has a contribution at the saddle point:
\bel
-S_{\text{FI}}=\int_{S^1 \times \D} \mathcal{L}_{\text{FI}}=2\pi  r \zeta \mathrm{Tr} i \beta r a.
\label{FIsaddle}
\ee

In the calculation of the localization, we deform the action by adding a $Q$-exact form $t Q\cdot V$ and take the limit $t  \to \infty$. 
In general, it is possible to add boundary $\mathcal{N}=(0,2)$ vector multiplets which are independent of 
the bulk $\mN=2$ vector multiplet, but we  treat boundary $\mathcal{N}=(0,2)$ vector multiplets which have originated from 
the  $\mN=2$ bulk vector multiplets.
Then
the 3d-2d coupled index on $S^1 \times \D$ can be evaluated by the one-loop calculation around the saddle point:
\bel
\mathcal{I}_{S^1 \times \D}&&=
\lim_{t_1,t_2 \to \infty } \int \mathcal{D} \Phi_{\text{3d}} \mathcal{D} \Phi_{\text{2d}} 
e^{-S[\Phi] -t_1 Q_{\text{3d}} \cdot V_{3d}[\Phi] -t_2 Q_{\text{2d}} \cdot V_{2d}[\Phi]}  \non
 &&=  \frac{1}{|W_G|} \int \frac{d^N (\beta r a)}{(2\pi)^N}  \left( \prod_{\alpha \neq 0} \sinh \frac{ i\beta r \alpha(a)}{2} \right)  e^{-S_{\text{cl}}} 
Z^{\text{3d}}_{1\text{-loop}} Z^{\text{2d}}_{1\text{-loop}} .
\ee
Here $\Phi_{\text{3d}}$ (resp.$ \, \Phi_{\text{2d}}$) is the collection of the 3d (resp.$\,$2d) fields in the models and  $S[\Phi]:= S_{\text{FI}}-S_{CS}-S_{b.CS}$. 
 $N$ is the rank of the gauge group $G$, 
$|W_G|$ is the order of the 
Weyl group $W_G$ of $G$ and $\alpha$ runs over the roots. 
 $\alpha (a) $ is the paring of a root $\alpha$ and a Cartan element $a$ in \eqref{saddlepoint}. 
 $S_{\text{cl}}$ is  the saddle point value of $S[\Phi]$.
We perform the  path integrals for the two dimensional boundary  fields except for  vector, chiral and Fermi multiplets 
which are obtained by the restriction of the three dimensional bulk vector and chiral multiplets.  
The factor ``$\sinh$" comes from the additional gauge fixing condition of $A_{\tau}$. 
For the closed manifolds $S^1 \times S^2$ and $S^3_b$, 
the Chern-Simons terms are $Q$-closed but not $Q$-exact and have contributions in the classical level.
On the other hand, in our case $S^1 \times \D$, we will see the Chern-Simons term contributes to  
one-loop determinants of the  boundary $\mathcal{N}=(0,2)$ (chiral and Fermi) multiplets through anomaly inflows.    
We summarize the results of the bulk and boundary one-loop determinants. The derivation is given in   appendix \ref{AppendixD}. 
\begin{itemize}
\item 
The one-loop determinant of 3d $\mN=2$ vector multiplet:

\bel
\left( \prod_{\alpha \neq 0} \sinh \frac{ i\beta r \alpha(a)}{2} \right) Z^{\text{3d.vec}}_{1\text{-loop}} =
 \prod_{\alpha\neq 0}  e^{\frac{-(i \beta r \alpha (a))^2}{8\beta_2}} (e^{i\beta r \alpha(a)};q^2)_{\infty}.
\label{oneloop3dV}
\ee 
Here we defined $q:=e^{-\beta_2}$, $(a;q)_{\infty}:=\prod_{n=0}^{\infty}(1-aq^n)$ and 
included the ``sinh" factor coming from the additional gauge fixing.

\item 
The one-loop determinant of 3d $\mN=2$ chiral multiplet with Neumann boundary condition:
\bel
Z^{\text{3d.chi.N}}_{1\text{-loop}} =  \prod_{\rho} 
e^{ \mathcal{E} (i\beta r \rho(a)+\Delta \beta_2+F_l M_l)}
(e^{-i\beta r \rho(a)-F_l M_l} q^{\Delta} ;q^2)^{-1}_{\infty}  ,
\label{oneloop3dCN}
\ee
with
\bel
\mathcal{E}(x):=\frac{\beta_2}{12} -\frac{1}{4} x + \frac{1}{8\beta_2} x^2.
\ee 
Here  
$\rho$ runs  over the weight of  the representation of the  gauge group  for the chiral multiplet.
The factor  $\mathcal{E}(x)$ comes from a zeta function regularization of zero point energies.

\item 
The one-loop determinant of 3d $\mN=2$ chiral multiplet with Dirichlet boundary condition:
\bel
Z^{\text{3d.chi.D}}_{1\text{-loop}} =  \prod_{\rho}  
e^{- \mathcal{E} ( -i\beta r \rho(a)+(2-\Delta) \beta_2-F_l M_l)}
(e^{i\beta r \rho(a)+F_l M_l} q^{2-\Delta} ;q^2)_{\infty}.
\label{oneloop3dCD}
\ee

\item 
The one-loop determinant of 2d $\mathcal{N}=(0,2)$ vector multiplet:
\bel
Z^{\text{2d}.\text{vec}}_{1\text{-loop}}
&&= \prod_{\alpha \neq 0} e^{-\frac{(i \beta r \alpha (a))^2}{4 \beta_2}} \th(e^{i \beta r \alpha(a)};q^2)  .
\label{oneloop2dV}
\ee

\item 
The one-loop determinant of 2d $\mathcal{N}=(0,2)$ chiral multiplet:
\bel
Z^{\text{2d.chi}}_{1\text{-loop}}= \prod_{\rho } 
e^{2\mathcal{E}( i\beta r \rho(a)+\Delta \beta_2+F_l M_l)}
 {\th(e^{- i \beta r \rho(a)- F_l M_l } q^{\Delta}; q^2 )}^{-1}_{\infty}.
\label{oneloop2dC}
\ee

\item 
The one-loop determinant of 2d $\mathcal{N}=(0,2)$ Fermi multiplet:
\bel
Z^{\text{2d.Fermi}}_{1\text{-loop}}= \prod_{\rho} 
e^{-2\mathcal{E}(i\beta r \rho(a)+\tilde{\Delta} \beta_2+F_a M_a)}
\th(e^{-i\beta r \rho(a)  -F_a M_a } q^{\tilde{\Delta}};q^2)_{\infty} .
\label{oneloop2dF}
\ee

\end{itemize}
We find that all the one-loop determinants do not depend on the fugacity $\beta_1$ explicitly.
Unless specifically mentioned, 
 we call the  one-loop determinants without the anomalous terms simply as the one-loop determinants.


Here it is worthwhile to make  several remarks on properties of these one-loop determinants. 
We shall recall there are two equivalent descriptions to express fluctuations 
along the normal directions to D-branes in the two dimensional hemisphere \cite{Honda:2013uca, Hori:2013ika}.
One is to impose the Dirichlet boundary condition for chiral multiplets whose lowest 
components label coordinates along the normal directions of D-branes. 
The other is to impose the Neumann boundary condition and introduce a boundary interaction 
which effectively transmutes the one-loop determinant with the Neumann boundary condition into 
that with the Dirichlet boundary condition.
Similarly, when we combine the three dimensional Neumann-type chiral multiplet and 
the boundary $\mathcal{N}=(0,2)$ Fermi multiplet with the common weight $\rho$ and R-charge $\tilde{\Delta}={\Delta}$, 
we can obtain a Dirichlet-type chiral multiplet in three dimensions:
\bel
Z^{\text{2d.Fermi}}_{1\text{-loop}} Z^{\text{3d.chi.N}}_{1\text{-loop}} =  Z^{\text{3d.chi.D}}_{1\text{-loop}}.
\ee 


When we consider 
the product of the Neumann-type and the Dirichlet-type chiral multiplets in three dimensions, 
we can construct a one-loop determinant of a chiral multiplet without magnetic charge $\mathbf{m}=0$ on $S^1 \times S^2$:
\bel
Z^{\text{3d.chi.N}}_{1\text{-loop}}  Z^{\text{3d.chi.D}}_{1\text{-loop}} 
=\frac{(e^{i\beta r \rho(a)+F_l M_l} q^{2-\Delta} ;q^2)_{\infty}}{(e^{-i\beta r \rho(a)-F_l M_l} q^{\Delta} ;q^2)_{\infty}}. 
\ee

Next we  comment on the two dimensional limit of these determinants.
 When the size of $S^1$ goes to zero $(\beta \sim \beta_2 \sim 0 ) $, 
the one-loop determinants of the three dimensional (vector and chiral) multiplets tend to the asymptotic forms 
up to exponential factors
 \bel
Z^{\text{3d}.\text{vec}}_{1\text{-loop}} && \sim \prod_{\alpha\neq 0}  \prod_{j=0}^{\infty}   (i r \alpha(a_{\text{2d}}) +j+1) ,\non
Z^{\text{3d.chi.N}}_{1\text{-loop}} &&\sim  \prod_{\rho}  \prod_{j=0}^{\infty} \left(-i r \rho(a_{\text{2d}})-F_l M_{\text{2d}. l}
 + \frac{{\Delta}}{2} +j \right)^{-1},\label{2dlimit}
\\ 
Z^{\text{3d.chi.D}}_{1\text{-loop}} &&\sim  \prod_{\rho}  \prod_{j=0}^{\infty} \left(i r \rho(a_{\text{2d}})+F_l M_{\text{2d}. l}
 -\frac{{\Delta}}{2} +1+j \right). \nonumber 
\ee
Here we defined $a_{\text{2d}}:=a \beta/2\beta_2$  and $M_{\text{2d}. l}:=M_l/2\beta_2$. 
These reproduce the  one-loop determinants of the $\mN=(2,2)$ vector and chiral multiplets on the  two dimensional hemisphere $\D$.  
In this limit, fugacities of the flavor symmetries become the twisted masses  in two dimensions.

\section{Relation to holomorphic blocks}
\label{Holomorphicblock}
In this section we study the relation between the 3d-2d coupled partition function $\mathcal{I}_{S^1 \times D^2}$ 
 and the holomorphic block $\mathcal{B}_{\alpha}$ in three dimensions \cite{Beem:2012mb}.
We  assume that  the  boundary multiplets couple to the  boundary gauge fields  which have originated in the bulk vector multiplets. 

First we consider a single chiral multiplet. Up to the anomalous contribution $\mathcal{E}$,
 the one-loop determinant with the Dirichlet boundary condition
 is expressed  as
\bel
Z^{\text{3d.chi.D}}_{1\text{-loop}}=(q^{2-\Delta} s^{-\rho} z^{-F_l}_{l};q^2)_{\infty} ,
\label{chiDwoan}
\ee 
where we defined $s^{\rho}:=e^{-i \beta r \rho (a)}$. 
\eqref{chiDwoan} is precisely the same formula as the contribution of the  chiral multiplet in the holomorphic block.

Next we discuss the condition to cancel the anomalous terms.
 As we have seen in (\ref{ClassicalCS}), there are anomalous contributions  which  break the single-valuedness under the large gauge transformation.  
 First let us study  anomalous terms for dynamical gauge fields, i.e., the quadratic  terms 
of the gauge fields around the saddle point.
The  condition for cancellation of the anomalous term  is given by
\bel
 -\kappa +\frac{\beta}{2\beta_2} \left( -I_2(\text{Ad}) +\sum_{i:\text{3d.chiral}} (-1)^{|i|} I_2(R_i) \right)=\frac{\beta}{\beta_2} \left (\sum_{n:\text{2d.Fermi}}  I_2(R_n)  -  \sum_{m:\text{2d.chiral}}  I_2(R_m) \right) ,\non
\label{anomalycan1} 
\ee
 with
 \bel
(-1)^{|i|}
=\left\{
\begin{array}{l}
 -1 \quad (\text{$i$-th chiral =Dirichlet}) \\
  +1 \quad (\text{$i$-th chiral =Neumann}).
\end{array}
\right.
\ee
Here $I_2(R)$ is the quadratic index of  a representation $R$ of   the Lie algebra of the dynamical gauge group and is defined by the relation 
$\mathrm{Tr}_{R}(T^a T^b)=I_2(R) \mathrm{Tr}_{\Box}(T^a T^b)$.  The symbol $\Box$ (resp. $\text{Ad}$  in \eqref{anomalycan1} ) denotes the fundamental (resp. adjoint) representation.

The  left hand side of (\ref{anomalycan1})  
is the contribution of the boundary  Chern-Simons term (\ref{ClassicalCS}), the 
one-loop anomalous term of the vector multiplet  (\ref{oneloop3dV}), and the one-loop anomalous terms 
of the chiral multiplets   (\ref{oneloop3dCN}),  (\ref{oneloop3dCD}), respectively.  The right hand side in (\ref{anomalycan1}) is the contribution from the anomalous terms $\mathcal{E}$'s of  the boundary multiplets (\ref{oneloop2dC}), (\ref{oneloop2dF}). The sums run over the  3d chiral multiplets, the 2d Fermi multiplets, and the 2d chiral multiplets, respectively. 
If we set $\beta_2=\beta$, 
the left hand side of \eqref{anomalycan1} reproduces the effective Chern-Simons level for the gauge group $G$,
 and the right hand side reproduces the  gauge anomaly coefficient of the boundary $\mathcal{N}=(0,2)$ theory.
Then (\ref{anomalycan1}) is the   anomaly inflow condition that describes the one-loop   
shift of the  classical Chern-Simons terms in  (\ref{treeanomaly}).  
A similar phenomenon is  known in four dimensional supersymmetric theory.
In the localization computation of the four dimensional  $\mathcal{N}=1$ superconformal index, there exist anomalous terms which are not invariant under large gauge transformations.
It was observed in \cite{Assel:2014paa} that the anomalous terms in the superconformal index are proportional to anomalies in four dimensions.

If there is a pair of a fundamental and an 
anti-fundamental chiral multiplets, we have an accidental cancellation. 
That is, we take a Neumann (Dirichlet) boundary condition 
for the fundamental (anti-fundamental) chiral multiplet respectively, then 
anomalous terms cancel out. Moreover  if we include an adjoint chiral multiplet, the contribution from  
the adjoint chiral multiplet cancels the anomalous factor coming from the vector multiplet. 
 This means that the Chern-Simons level on $S^1 \times D^2$ cannot be 
shifted in $\mathcal{N}( \ge 3) $ supersymmetric cases, and 
the condition (\ref{anomalycan1}) for $\mathcal{N} \ge 3$ cases matches with (\ref{treeanomaly}).    
 This observation is similar to  the level shifts on the flat space \cite{Kao:1995gf}.

 Next we  consider  the cancellation of anomalous terms for the mixed Chern-Simons terms between 
 the $a$-th central $U(1)$ gauge symmetry  and the $l$-th $U(1)$ flavor symmetry: 
 \bel
 &&-\kappa_{a \, l}  +\frac{\beta}{2 \beta_2} \sum_{i:\text{3d.chiral}} (-1)^{|i|}  Q^i_a F^i_l \non
&&~~~~~~~~~~~~~~~~~=\frac{ \beta }{ \beta_2} \left( \sum_{n:\text{2d.Fermi}} \tilde{Q}^n_{\text{2d}, a} 
 \tilde{F}^n_{\text{2d}, l}  -  \sum_{m:\text{2d.chiral}} Q^m_{\text{2d}, a} F^m_{\text{2d}, l} \right). 
\label{anomalycan2}
 \ee
 Here we used the fact that 
the background gauge field for the $l$-th flavor symmetry is $A^{(l)}_{\mu}=\left(0, 0, -i M_l / \beta r \right)$. 
$Q^i_a$ is the $a$-th central $U(1)$ gauge charge for the $i$-th 3d chiral multiplet. 
 $\tilde{Q}^n_{\text{2d}, a}$ (resp. $Q^m_{\text{2d}, a}$) denotes the 
$a$-th central $U(1)$ gauge charge for two dimensional Fermi (resp. chiral) multiplet.  
 $\tilde{F}^n_{\text{2d}, l}$ (resp. $F^m_{\text{2d}, l}$) denotes the 
$l$-th $U(1)$ flavor charge for the two dimensional $n$-th Fermi (resp. $m$-th chiral) multiplet.  

In addition to the gauge and flavor mixed CS terms, we have the R-symmetry mixed Chern-Simons levels $\kappa_{s \, R}$.
Here the subscript ``$R$'' means the R-symmetry and ``$s$" runs over the center of the gauge symmetry and  the Cartan of flavor symmetries. 
Then the mixed Chern-Simons terms are given by
\bel
S^{sR}_{CS}=\frac{i\kappa_{s \, R}}{4\pi} \int_{S^1 \times D^2} A^{(s)} d A^{(R)} .
\ee
Here $A^{(R)}_{\mu}$ is the background gauge field coupled to the R-symmetry current. 
As in the cases of the gauge-gauge or gauge-flavor mixed boundary Chern-Simons terms (\ref{quivermixedbt}), 
these boundary terms are expected to contain
quadratic terms of the gauge field:
\bel
\int_{T^2} A^{(s)}_{z} A^{(R)}_{\bar{z}} .
\ee  
Since the twisted  boundary condition \eqref{eq:SUSYind} on $S^1 \times \D$ is the same as the condition    \eqref{eq:SUSYind0} on $S^1 \times S^2$, 
the background gauge field $A^{(R)}_{\mu}$ on  $S^1 \times D^2$ is given by  the one on $S^1 \times S^2$.
When we put $\beta_2 =\beta$, the background gauge field  \cite{Closset:2012ru} is written as
\bel
A^{(R)}_{\mu}=\left(0,0,-\frac{i}{r} \right).
\ee
Then the condition for the cancellation of anomalous terms including   the gauge and R-symmetries is 
given by 
 \bel
 &&- \kappa_{a R}  + \sum_{i:\text{3d.chiral}} \frac{(-1)^{|i|}}{2}  Q^i_a (\Delta_i-1) \non
&& ~~~~~~~~~~~~~~~~~~ =\sum_{n:\text{2d.Fermi}}  \tilde{Q}^n_{\text{2d}, a}  (\tilde{\Delta}_n-1) 
 -  \sum_{m:\text{2d.chiral}}  Q^m_{\text{2d}, a}(\Delta_m-1),
\label{anomalycan3}
 \ee
and that for the  flavor and R-symmetries is given by
 \bel
 &&-  \kappa_{l R}  + \sum_{i:\text{3d.chiral}} \frac{(-1)^{|i|}}{2} F^{i}_l (\Delta_i-1) \non
&& ~~~~~~~~~~~~~~~~~~ =\sum_{n:\text{2d.Fermi}}  \tilde{F}^n_{\text{2d}, l} (\tilde{\Delta}_n-1) 
 -  \sum_{m:\text{2d.chiral}}  F^m_{\text{2d}, l} (\Delta_m-1).
\label{anomalycan4} 
\ee

When we take  Abelian gauge groups  and put $\beta_2 =\beta$, 
the  conditions (\ref{anomalycan1}), (\ref{anomalycan2}), (\ref{anomalycan3}), (\ref{anomalycan4}) 
agree with the decomposition rule for the effective mixed Chern-Simons levels proposed in the 
holomorphic blocks \cite{Beem:2012mb}.
Let us study several examples  of 3d-2d indices and compare holomorphic blocks with our 3d-2d indices in the following subsections.
We take mixed bare Chern-Simons levels and charges for 
gauge, flavor and R-symmetries as in \cite{Beem:2012mb}. 
Since  2d multiplets are chosen to cancel the anomalous terms of 3d bulk theory,  
we call the  one-loop determinants without the anomalous terms simply as the one-loop determinants.
 
\subsection{Mirror of  $T_{\Delta}$}
As a first example, we take $G=U(1)_G$ with a bare dynamical Chern-Simons level $k_{G G}=+\frac{1}{2}$ and consider 
a single chiral multiplet with a charge $+1$. This model is the 3d mirror dual of the tetrahedron $T_{\Delta}$. 
The charge assignments and the effective mixed Chern-Simons levels \footnote{The effective Chern-Simons level is related to 
the bare Chern-Simons level through an equation;
 $\kappa^{\text{eff}}_{s t}=\kappa^{\text{bare}}_{s t}
+\sum_{l} \frac{1}{2}\text{sign} (m_l) Q^{l}_s Q^{l}_t  $. In \cite{Beem:2012mb},  signatures of 
fermion masses are taken as $\text{sign} (m_l)=+1$. } are listed in Table \ref{chargetetra}.
\begin{table}[htb]
\begin{center}
\begin{tabular}{|c||ccc|}
\hline
  &   $U(1)_G$ & $U(1)_J$  & $U(1)_R$ \\
\hline
 $\phi$  &   $1$ & $0$ & $0$  \\
\hline
\end{tabular} 
\hspace{2.0cm}
\begin{tabular}{|c|ccc|}
\hline
 $\kappa^{\text{eff}}_{st}$ &   $U(1)_G$ & $U(1)_J$  & $U(1)_R$ \\
\hline
 $U(1)_G$   & $1$ & $1$ & $-1$  \\
$U(1)_J$    & $1$ & $0$ & $0$ \\
 $U(1)_R$   & $-1$ & $0$ & $0$ \\
\hline
\end{tabular}
\end{center}
\caption{Left: the charge assignments of the  scalar in the  
3d chiral multiplet. $U(1)_G$ is the dynamical gauge group and $U(1)_J$ is the topological 
flavor group. Right: the set of mixed Chern-Simons levels. }
\label{chargetetra}
\end{table} 
Then the one-loop determinant of the 3d bulk chiral multiplet is given by
\bel
Z^{\text{3d.chi.D}}_{1\text{-loop}} =(q^2 s^{-1};q^2)_{\infty},
\label{loopvor}
\ee 
where $s:=e^{-i \beta r a}$. In oder to satisfy
 the conditions (\ref{anomalycan1}), (\ref{anomalycan2}), (\ref{anomalycan3}), (\ref{anomalycan4}), we introduce a  
 boundary chiral multiplet 
and a Fermi multiplet, 
whose lowest components are  denoted by a scalar $\phi'$ 
and a fermion  $\Psi$, respectively.
 The  charge assignments of these boundary multiplets are listed in Table \ref{vorboundary}
\begin{table}[htb]
\begin{center}
\begin{tabular}{|c||ccc|}
\hline
  &   $U(1)_G$ & $U(1)_J$  & $U(1)_R$ \\
\hline
 $\phi' $  &   $1$ & $1$ & $0$  \\
 $\Psi$  &   $0$ & $1$ & $-1$  \\
\hline
\end{tabular} 
\end{center}
\caption{The charge assignments of the scalar in the 2d boundary chiral multiplet and the fermion in the Fermi multiplet. }
\label{vorboundary}
\end{table} 
and the one-loop contribution of the boundary multiplets is given by
\bel
Z^{2d}_{1\text{-loop}}=\frac{\th (x;q^2)}{\th (s x;q^2)}.
\label{loopbvor}
\ee
From (\ref{loopvor}) and (\ref{loopbvor}), we obtain a 3d-2d index  of this model ${T}^{\prime}_{\Delta}$
mirror dual to the  tetrahedron theory ${T}_{\Delta}$:
\bel
\mathcal{I}^{{T^{\prime}}_{\Delta} }_{S^1 \times \D}=\int \frac{d s}{2\pi i s} \frac{\th (x;q^2)}{\th (s x;q^2)} (q^2 s^{-1};q^2)_{\infty}. 
\label{tetraD}
\ee 
\eqref{tetraD}  is exactly the same as  the holomorphic block of ${T^{\prime}}_{\Delta}$.\footnote{ The convention of the theta function 
is $\th(x;q) (\text{here})=\th(q^{\frac{1}{2}}x;q)$(\cite{Beem:2012mb}). 
The normalization of the fugacity $q$ is different 
from that of \cite{Beem:2012mb}. In addition, 
an extra sign difference comes from  $(-1)^{R}$ which is used in the holomorphic blocks instead of $(-1)^{F}$. 
Then the identification becomes $q (\text{here}) =-q^{\frac{1}{2}} $(\cite{Beem:2012mb}).  }
That is to say, in our language, the contribution of the chiral multiplet in the holomorphic blocks \cite{Beem:2012mb} corresponds to the one-loop determinant of three dimensional chiral multiplets with the {\em Dirichlet boundary condition}.  

Next we consider the Neumann boundary condition and study the relation between our 3d-2d index and the holomorphic blocks.
At least when the gauge group is Abelian and the superpotential is absent,
we can reproduce the homolorphic blocks from the  3d-2d  index.
To see this, we take the Neumann boundary condition for the chiral multiplet in ${T}^{\prime}_{\Delta}$.
The one-loop determinant of three dimensional chiral multiplet is given by
\bel
Z^{\text{3d.chi.N}}_{1\text{-loop}} =( s;q^2)^{-1}_{\infty}.
\ee
In this case, the   level shift from the 3d anomalous factor has  the opposite  sign 
of  the Dirichlet boundary condition.
Then  the dynamical Chern-Simons level is shifted  by $-\frac{1}{2}$.
The effective Chern-Simons levels for the Neumann boundary condition are listed in Table \ref{chargetetra2}.
In addition, we have to satisfy 
the conditions (\ref{anomalycan1}), (\ref{anomalycan2}), (\ref{anomalycan3}), (\ref{anomalycan4}) 
for the anomaly cancellation and introduce appropriate 
 boundary multiplets. The charge assignments of the boundary multiplets are
  listed in Table \ref{vorboundary2}. Then the one-loop determinant of the boundary multiplets is given by
\bel
Z^{\text{2d}}_{1\text{-loop}}=\frac{\th (s;q^2) \th (x;q^2)}{\th (sx;q^2)}.
\ee 
The 3d-2d index with the Neumann boundary condition 
\bel
\mathcal{I}^{{T'}_{\Delta}. \text{Neu} }_{S^1 \times \D}=\int \frac{d s}{2\pi i s} ( s;q^2)^{-1}_{\infty}  \frac{\th (s;q^2) \th (x;q^2)}{\th (sx;q^2)}
\ee
gives the same result as the 3d-2d index with the Dirichlet boundary condition (\ref{tetraD}).

\begin{table}[htb]
\begin{center}
\begin{tabular}{|c|ccc|}
\hline
 $\kappa^{\text{eff}}_{st}$ &   $U(1)_G$ & $U(1)_J$  & $U(1)_R$ \\
\hline
 $U(1)_G$   & $0$ & $1$ & $0$  \\
$U(1)_J$    & $1$ & $0$ & $0$ \\
 $U(1)_R$   & $0$ & $0$ & $0$ \\
\hline
\end{tabular}
\end{center}
\caption{The set of mixed effective  Chern-Simons levels for the Neumann boundary condition.}
\label{chargetetra2}
\end{table}

\begin{table}[htb]
\begin{center}
\begin{tabular}{|c||ccc|}
\hline
  &   $U(1)_G$ & $U(1)_J$  & $U(1)_R$ \\
\hline
 $\phi'' $  &   $1$ & $1$ & $0$  \\
 $\Psi_1$  &   $1$ & $0$ &   $-1$  \\
 $\Psi_2$  &   $0$ & $1$ & $-1$  \\
\hline
\end{tabular} 
\end{center}
\caption{The charge assignments of the scalar in the 2d boundary chiral multiplet and fermions in the Fermi multiplets. }
\label{vorboundary2}
\end{table}

\subsection{XYZ model}
\label{XYZmodel}

As a second example, we consider the XYZ model that consists of three chiral multiplets  in bulk three dimensions (Table \ref{XYZcontent}).
\begin{table}[htb]
\begin{center}
\begin{tabular}{|c||ccc|}
\hline
  &   $U(1)_x$ & $U(1)_y$  & $U(1)_R$ \\
\hline
 $\phi_1$  &   $1$ & $0$ & $0$  \\
$\phi_2$  &   $0$ & $1$ & $0$ \\
 $\phi_3$ &  $-1$ & $-1$ & $2$ \\
\hline
\end{tabular} 
\hspace{2.0cm}
\begin{tabular}{|c|ccc|}
\hline
 $\kappa^{\text{eff}}_{st}$ &   $U(1)_x$ & $U(1)_y$  & $U(1)_R$ \\
\hline
 $U(1)_x$   & $1$ & $1$ & $-1$  \\
$U(1)_y$    & $1$ & $1$ & $-1$ \\
 $U(1)_R$   & $-1$ & $-1$ & $1$ \\
\hline
\end{tabular}
\end{center}
\caption{Left: the charge assignments of scalars in the  3d chiral multiplets.  Right: the set of mixed Chern-Simons levels}
\label{XYZcontent}
\end{table} 
When we impose  the Dirichlet boundary conditions on the bulk chiral multiplets,  
the one-loop determinant is given by
\bel
Z^{\text{3d.chi}}_{1\text{-loop}}=(q^2 x^{-1}; q^2)_{\infty} (q^2 y^{-1}; q^2)_{\infty} ( x y; q^2)_{\infty}.
\label{XYZ1loop}
\ee

\begin{table}[htb]
\begin{center}
\begin{tabular}{|c||ccc|}
\hline
  &   $U(1)_x$ & $U(1)_y$  & $U(1)_R$ \\
\hline
 $\phi$  &   $1$ & $1$ & $0$  \\
\hline
\end{tabular} 
\end{center}
\caption{The charge assignment of the scalar in the 2d boundary chiral multiplet. }
\label{table:XYZboundary}
\end{table} 
Next we introduce  a boundary  chiral multiplet $\phi$ to cancel the bulk-boundary anomalies (Table \ref{table:XYZboundary}). 
Then this multiplet has the contribution at the one-loop level:
\bel
Z^{\text{2d.chi}}_{1\text{-loop}}=\th(x y ;q^2)^{-1}.
\label{bXYZ1loop}
\ee
From (\ref{XYZ1loop}) and (\ref{bXYZ1loop}), the 3d-2d index on $S^1 \times \D$
becomes
\bel
\mathcal{I}^{XYZ}_{S^1 \times \D}=\frac{(q^2 x^{-1}; q^2)_{\infty} (q^2 y^{-1}; q^2)_{\infty} ( x y; q^2)_{\infty}}{\th(x y ;q^2)} . 
\ee
This matches with the holomorphic block of the XYZ model.

\subsection{SQED}

\begin{table}[htb]
\begin{center}
\begin{tabular}{|c||cccc|}
\hline
 &   $U(1)_G$ &   $U(1)_x$ & $U(1)_y$  & $U(1)_R$ \\
\hline
 $\phi_1$  &   $1$ &   $0$ & $0$ & $0$  \\
$\phi_2$  &   $-1$ &   $1$ & $0$ & $0$ \\
 \hline
\end{tabular} 
\hspace{1.0cm}
\begin{tabular}{|c|cccc|}
\hline
 $\kappa^{\text{eff}}_{st}$ &   $U(1)_G$ &   $U(1)_x$ & $U(1)_y$  & $U(1)_R$ \\
\hline
$U(1)_G$  & $1$  & $0$ & $1$ & $-1$  \\
 $U(1)_x$  & $0$  & $0$ & $0$ & $0$  \\
$U(1)_y$   & $1$  & $0$ & $0$ & $0$ \\
 $U(1)_R$  & $-1$  & $0$ & $0$ & $0$ \\
\hline
\end{tabular}
\end{center}
\caption{Left: the charge assignments of scalars in the 3d chiral multiplets. Right: 
the set of effective mixed Chern-Simons levels}
\label{SQEDcontent}
\end{table} 
In this subsection, we consider the SQED model. 
From Table \ref{SQEDcontent}, the one-loop determinant of three dimensional chiral multiplets $\phi_1, \phi_2$ is given by
\bel
Z^{\text{3d.chi.D}}_{1\text{-loop}}=(s^{-1}  q^2 ;q^2)_{\infty} (s q^2 x^{-1} ;q^2)_{\infty}.
\ee 
A pair of  boundary multiplets should be introduced  to cancel anomalous terms. 
It is  a pair of a 
 boundary $\mathcal{N}=(0,2)$ chiral  
and  a Fermi multiplets whose lowest components are respectively a 
scalar $\phi'$  and a fermion $\Psi$.
Their charge assignments are listed in Table \ref{SQEDboundary}. 
Then the one-loop contributions of the boundary multiplets are given by
\bel
Z^{\text{2d.chi}}_{1\text{-loop}} Z^{\text{2d.Fermi}}_{1\text{-loop}}=\frac{\th (y;q^2)}{\th (s y;q^2)}.
\ee 

\begin{table}[htb]
\begin{center}
\begin{tabular}{|c||cccc|}
\hline
 &   $U(1)_G$  &   $U(1)_x$ & $U(1)_y$  & $U(1)_R$ \\
\hline
  $\phi'$  & $1$  &   $0$ & $1$ & $0$  \\
   $\Psi$  & $0$  &   $0$ & $1$ & $0$  \\
\hline
\end{tabular} 
\end{center}
\caption{The charge assignment of the scalar (resp. fermion) in the boundary 
chiral (resp. Fermi) multiplet. }
\label{SQEDboundary}
\end{table} 
Thus the 3d-2d index of this model becomes 
\bel
\mathcal{I}^{\text{SQED}}_{S^1 \times D^2}=\int \frac{d s}{2\pi i s } (s^{-1}  q^2 ;q^2)_{\infty} (s q^2 x^{-1} ;q^2)_{\infty} \frac{\th (y;q^2)}{\th (s y;q^2)}.
\ee
This has the same expression as the result in the holomorphic block for the SQED. 
The holomorphic block for the SQED also matches with that in the XYZ model. Thus 3d-2d indices for these two
 models produce the identical result. 
We make a comment here; the SQED and the XYZ model flow to the same IR fixed point  and  
a pair of these models  is the simplest example of the $\mathcal{N}=2$ mirror symmetry in three dimensions \cite{Aharony:1997bx}.
The half BPS boundary conditions for the SQED and the XYZ model were studied in \cite{Okazaki:2013kaa}, where
it is shown that  the $\mathcal{N}=(0,2)$-type BPS boundary condition in the SQED  
is mapped to the $\mathcal{N}=(0,2)$-type supersymmetry in the XYZ model.
 Our result is  consistent with their analysis of the boundary conditions 
because the 3d-2d index on $S^1 \times D^2$ 
preserve the boundary $\mN=(0,2)$ supersymmetry. 
This situation is different from the two dimensions. For a mirror pair in two dimensions, 
the A-type boundary supersymmetry  is mapped to the B-type boundary  supersymmetry  \cite{Ooguri:1996ck, Hori:2000kt, Hori:2000ck}.    
\subsubsection*{A remark on non-Abelian gauge theories}
For non-Abelian gauge theories, 
there is   a difference  between the one-loop determinant of  the vector multiplet \eqref{oneloop3dV} 
and 
the contribution of the vector multiplet in the holomorphic blocks. 
In our calculation, up to the anomalous term,  the contribution of the vector multiplet is given by the one-loop determinant:
\bel
{\prod_{\alpha > 0}( s^{\alpha};q^2)_{\infty} ( s^{-\alpha};q^2)_{\infty}} .
\label{vecminimal}
\ee
On the other hand, the holomorphic block in \cite{Beem:2012mb} leads to 
the contribution of the vector multiplet with a non-Abelian gauge group 
\bel
\prod_{\alpha >0} \frac{(q s^{\alpha};q^2)_{\infty} (q^{-1} s^{-\alpha};q^2)_{\infty}}{(q^{2} s^{\alpha};q^2)_{\infty} (q^{-2} s^{-\alpha};q^2)_{\infty}}.
\label{vecholomorphid}
\ee   
In general, we expect that our indices for non-Abelian gauge theories are
 different from the holomorphic blocks for non-Abelian gauge theories. 
We will see later  that our index for 
3d $\mathcal{N}=4$ $U(N)$ gauge theory without boundary degrees of freedom 
 agrees with an index on $S^1 \times \mathbb{C}$ proposed in \cite{Aganagic:2013tta} .

\section{Several models}
\label{models}
In this section   we study  properties of indices in  several examples.
Unless specifically mentioned, 
we choose the bare Chern-Simons levels to cancel the anomalous terms coming from the 
one-loop determinants.

\subsection{$q$-deformed Whittaker function and K-theoretic $J$-function for $\C \mathbf{P}^{N}$}
In this subsection we explain  mathematical aspects of  the index of  $G=U(1)$ with $N+1$ chiral multiplets with 
gauge charges $+1$'s:
\bel
\mathcal{I}^{\C \mathbf{P}^{N}}_{S^1 \times \D}&=&\int \frac{d s}{2 \pi i s } 
  \frac{s^{-2\pi r \zeta} }{\prod_{l=1}^{N+1} \prod_{j=0}^{\infty} (1-s  q^{2j} z_l)} .
\label{bulkU1}
\ee
Here $s=e^{-i\beta r a}$, $q=e^{- \beta_2}$, $z_l=e^{-M_l}$ and $M_l$'s $(l=1, \cdots, N+1)$ represent 
the set of fugacities of $SU(N+1)$ flavor symmetry.

First let us recall the mathematical  aspects of the hemisphere partition function.  
In two dimensions, the partition function on the hemisphere $D^2$ is 
 related to  the (equivariant) $J$-function of the $\C \mathbf{P}^{N}$ 
model in the large volume regime.
In order to clarify the geometric data in our model, we evaluate the above integral explicitly.
The integrand has poles and 
we consider residues at $s=q^{-2k}   z^{-1}_{l'}, (k=0,1 \cdots)$.
Then (\ref{bulkU1}) is rewritten as
\bel
\mathcal{I}^{\C \mathbf{P}^N}_{S^1 \times \D}&=&\oint_{ s =z^{-1}_{l'}} \frac{d s}{2 \pi i s }  z^{2\pi r \zeta }_{l'} s^{-2\pi r \zeta} 
 \Bigl(  \frac{ 1}
{\prod_{l=1}^{N+1} \prod_{j=0}^{\infty} (1- s z_l q^{2j})}  \Bigr)
\non
&& ~~~~~~~~~~~~~~~~ \qquad  \times \left( \sum_{k=0}^{\infty} 
 \frac{         Q^{k} }
{  \prod_{l=1 }^{N+1} \prod_{j=1 }^{k}  (1- s z_{l}  q^{-2j}  )}  \right).
\label{bulkU2}
\ee
Here we defined $Q:=q^{2\pi r \zeta}$ and assumed $\zeta > 0$. The region for $\zeta > 0$  corresponds to 
the Higgs branch in the two dimensional limit.
The second line in \eqref{bulkU2}
\bel
J^{\C \mathbf{P}^{N}}(Q,s,z,q):=\sum_{k=0}^{\infty} 
 \frac{        Q^{k}     }
{  \prod_{l=1 }^{N+1} \prod_{j=1 }^{k}  (1- s z_{l}  q^{-2j}  )}
\ee
agrees with the equivariant K-theoretic $J$-function of $\C \mathbf{P}^{N}$  \cite{Givental:2001} by  rescaling parameters  appropriately. 
This function $J^{\C \mathbf{P}^{N}}$  
is reduced to the ordinary K-theoretic $J$-function in the  limit $z_{l}\rightarrow 1$. 
In order to compare our model to two dimensional cases, we take the two dimensional limit (\ref{2dlimit}).
Then the index (\ref{bulkU1}) is reduced to the hemisphere 
partition function of the 2d model where the moduli space of  the Higgs branch  is $\C \mathbf{P}^{N}$:
\bel
\lim_{\beta \to 0} \mathcal{I}_{S^1 \times \D} \sim \int \frac{d y}{2 \pi i  } 
  e^{2\pi i \zeta_{2d} y}{\prod_{l=1}^{N+1}  \Gamma (y-M_l)} .
\ee
Namely, this has the same formula as the two dimensional hemisphere partition function for the 
$\mN=(2,2)$ $U(1)$ theory with
$N+1$ chiral multiplets with 
gauge charges $+1$'s and twisted masses $M_l$'s.

In \cite{Gerasimov1, Gerasimov2}, an eigenfunction of the Hamiltonian of the $q$-deformed $\mathfrak{gl}_{N+1}$-Toda chain is constructed, that is, 
 the
 $q$-deformed  Whittaker function.
This 
 $q$-deformed Whittaker function $\Psi^{\mathfrak{gl}_{N+1}}_{z_i}(n,k)$ has a following contour integral representation 
\bel
\Psi^{\mathfrak{gl}_{N+1}}_{\mathbf{z}}(n,k)=\left( \prod_{l=1}^{N} z^{k}_{l} \right) \oint \frac{ds }{2\pi i s} s^{-n} \prod_{l=1}^{N+1} (z_l s;q)^{-1}_{\infty}.
\ee
When we set $n=2\pi r \zeta$ and replace $q \to q^2$ in the above equation,  the index 
{(\ref{bulkU1})} on $S^1 \times D^2$ agrees with 
this 
$q$-deformed Whittaker function $\Psi^{\mathfrak{gl}_{N+1}}_{\mathbf{z}}(n, k)$ up to an overall constant $\left( \prod_{l=1}^{N} z^{k}_{l} \right)$. 
Here the contour is chosen to enclose all the poles except for the pole at the origin.
We can also include the factor $\left( \prod_{l=1}^{N} z^{k}_{l} \right)$ in
the index on $S^1 \times D^2$ by turning on the FI term (\ref{FIterm}) for the flavor gauge field with the 
FI-parameter $\frac{k}{2\pi i}$. 
The geometrical interpretation of  $\Psi^{\mathfrak{gl}_{M}}_{\mathbf{z}}(n,k)$ was conjectured in
\cite{ Gerasimov2} as
\bel
\left( \prod_{l=1}^{N} z^{k}_{l} \right) \mathcal{I}^{\C \mathbf{P}^N}_{S^1 \times \D} 
= \int_{ \mathcal{Q M}_{\infty} (\C \mathbf{P}^{N})}
\text{Ch}_{G} ( \mathcal{L}_k \otimes \mathcal{O} (n) ) \text{Td}_G (T \mathcal{Q M}_{\infty} (\C \mathbf{P}^{N})) .
\ee
Here $\mathcal{Q M}_{\infty} (\C \mathbf{P}^{N})$ is  the space of the  degree-$\infty$   quasimaps $\C \mathbf{P}^1 \to \C \mathbf{P}^N$ 
and  $\text{Ch}_{G}$ and  $\text{Td}_G$ are $G=U(1) \times GL(N+1)$-equivariant Chern character and Todd class, respectively.

Remarkably, it was also conjectured in \cite{Gerasimov4} that the $q$-deformed Whittaker function is related to    
the partition function of an equivariant A-type twisted model on $S^1 \times \D$. 
Three dimensional version of the A-type twisted Chern-Simon-Matter theory is also constructed in \cite{Ohta:2012ev}. 
But the supersymmetry considered in this paper is not topologically twisted supersymmetry.
It would be interesting to reveal the precise relation among the 3d-2d index studied in this paper, quasimaps, and the A-type twisted theories in three dimensions.

\subsection{Vortex partition function and surface operator}
\label{vpf}
The vortex partition functions \cite{Shadchin:2006yz, Dimofte:2010tz, Yoshida:2011au} are vortex counterparts of the Nekrasov instanton partition functions \cite{Nekrasov:2002qd}.
It was shown in \cite{Pasquetti:2011fj, Hwang:2012jh, Taki:2013opa}
 that the partition function on $S^3_b$ ($S^1 \times S^2$) is respectively
 factorized into a pair of the vortex and the anti-vortex partition functions.
From the view point of the Higgs branch localization in three dimensions \cite{Fujitsuka:2013fga, Benini:2013yva}, 
we can also construct some $Q$-exact term whose 
saddle points admit point-like vortices  at the north pole of the hemisphere. 
Thus we expect the index on $S^1 \times D^2$ contains contributions from vortex partition functions.

Here we consider a $U(N)$ SQCD; the gauge group is $G=U(N)$ and 
the flavor symmetry is $SU(N_f)\times SU(\tilde{N}_f)$ with $N_f \ge \tilde{N}_f$.
We have $N_f$  fundamental  chiral  multiplets with 
the Neumann boundary conditions and $\tilde{N}_f$  
anti-fundamental  chiral  multiplets  with the Dirichlet boundary conditions.
The index of this model  is given by
\bel
\mathcal{I}_{S^1 \times \D}&=&\frac{1}{N!}\int \prod_{a=1}^{N} \frac{d s_a}{2 \pi i s_a } s^{-2\pi r \zeta  }_a \prod_{1 \le a \neq b  \le N } 
 \prod_{j=0}^{\infty}  (1- s_a s^{-1}_b q^{2j})   
 \prod_{a=1 }^{N}  \frac{\prod_{m=1}^{\tilde{N}_f} (1-s_a q^{2j+2}  \tilde{z}^{-1}_m) }{\prod_{l=1}^{N_f} (1-s_a  q^{2j} z_l)} .\non
\ee
Here $s_b=e^{-i\beta r a_b}$, $q=e^{- \beta_2}$, $z_l=e^{-M_l}$ and $M_l$'s $(l=1, \cdots, N_f)$ represent 
the set of fugacities of the $SU(N_f)$ flavor symmetry. Also
we put $\tilde{z}_m:=e^{-\tilde{M}_m}$ and $\tilde{M}_m$'s $ (m=1, \cdots, \tilde{N}_f)$ denote 
the set of fugacities of the flavor symmetry $SU(\tilde{N}_f)$ .

When we take residues at poles $s_a=q^{-2{j'}_a}  z^{-1}_{{l'}_a}$ 
 $({l'}_a=1, \cdots, N_f ;a=1, \cdots, N; {j'}_a=0,1,2,\cdots)$, 
then the partition function is written as a combination of 
classical terms, one-loop terms and vortex partition functions
\bel
&&\mathcal{I}_{S^1 \times \D} =\sum_{\{ l' \} \subset \{ N_f\}}Z^{ \{ l' \}}_{\text{cl}} Z^{\{ l' \}}_{1\text{-loop}} Z^{\text{vortex}, \{ l' \} }_{\text{K-theory}},
\label{factorizationUN}
\ee
with
\bel
&&Z^{ \{ l' \}}_{\text{cl}}=\prod_{a=1}^N (z^{-1}_{{l'}_a} )^{2\pi r \zeta}, \\
&&Z^{\{ l' \}}_{1\text{-loop}} =\prod_{j=0}^{\infty}  \frac{   \prod_{m=1}^{\tilde{N}_f}(1- q^{2j+2} {z}^{-1}_{{l'}_{a}} \tilde{z}^{-1}_m)  }{ 
 (1- q^{2j+2})^{N} \prod_{a=1}^{N} \prod_{l \notin \{ {l'}\}}   (1- z_l z^{-1}_{{l'}_{a}}q^{2j})^{-1}}, \\
&&Z^{\text{vortex}, \{ l'\}}_{\text{K-theory}}= \sum_{ \{ k \}}Q^{\sum_{a=1}^N{k}_a} \non
&& \times \frac{ \prod_{j=1}^{{k}_{a}} \prod_{s=1}^{\tilde{N}_f}(1- q^{-2j+2} {z}^{-1}_{{l'}_{a}} \tilde{z}^{-1}_s)}
{  \Bigl(  \prod_{1  \le a, b \le N}  \prod_{j=0}^{{k}_b-1} (1- z_{{l'}_b} z^{-1}_{{l'}_{a}}q^{2j-2{k}_{a}})\Bigr) 
\Bigl( \prod_{a=1}^{N} \prod_{l \notin \{ {l'}\}} \prod_{j=1}^{{k}_a}  (1- z_l z^{-1}_{{l'}_{a}}q^{-2j}) \Bigr)}. \non
\label{ktheoreticvortex}
\ee
Here we defined $\{ l'\}:=\{l_1, l_2, \cdots, l_N \}$ with $1\le l_1 < l_2 < \cdots < l_N \le N_f$
and $\{ N_f \}:=\{1, 2 \cdots, N_f \}$. 
Also the sum is defined by $\sum_{ \{ j'\}}:=\sum_{i=1}^N \sum_{j_i=0}^{\infty} $.
In three dimensions, BPS vortices are particle-like object and the 
K-theoretic  vortex partition functions contribute to the BPS index. 
The appearance of the K-theoretic  vortex partition functions in (\ref{factorizationUN})  
is also  consistent with the observation that the index on $S^1 \times \D$ is related to 
the 3d $\mN=2$ BPS index on $S^1 \times \R^2$. 
This is  analogous to the fact that  instantons on $S^1 \times \R^4$ are particle-like objects and 
the K-theoretic instanton partition functions \cite{Nakajima:2005fg} contribute to the BPS index in five dimensions.   

Next, we study the vortex partition function \eqref{ktheoreticvortex} from the view point of  the instanton counting with surface operators.
In   the geometric engineering, 
 instanton counting with surface operators in five dimensions is expected to be encoded in partition functions 
of  open-closed (refined) topological strings.   Vortex partition functions of $U(1)$ gauge theories arise 
in a certain limit of   topological string partition functions 
 \cite{Dimofte:2010tz}.  The correspondence between the vortex partition functions and the  topological strings  was extended to  non-Abelian gauge theories \cite{Taki:2013opa}. 
On the other hand,   it has not been studied how 
vortex partition functions of non-Abelian gauge theories appear in the instanton partition functions with  surface operators.
We will  show that the vortex counting for the non-Abelian $U(N)$ gauge theory with $N_f$ fundamental chiral multiplets arise  
in the sector of vanishing  instanton number.

We consider a five dimensional $\mathcal{N}=1$ pure $SU(N_f)$ gauge theory  on $S^1 \times \C^2$. 
We  take the  surface operator specified by a Levi subgroup $\mathbb{L}=S (U(N) \times U(N_f -N) ) \subset SU(N_f)$.  
Then  the instanton counting with the surface operator is replaced by the instanton counting on the orbifold $\C \times \C/\Z_2$ \cite{Kanno:2011fw, Kanno:2012}. 
By the orbifold action $(z, \omega ) \to (z, - \omega)$, the set of the 
ADHM data $(\bm{ B}_1, {\bm B}_2, {\bm I}, {\bm J})$:
\bel
{\bm B}_1, {\bm B}_2 \in \mathrm{End} (V), \quad {\bm I} \in \text{Hom} (W, V), \quad 
{\bm J} \in \text{Hom} (V, W)  
\ee
is divided into  two groups  with $\Z_2$-grading indices ``$i$" ($i=1,2$) 
\bel
V=V_1 \oplus V_2, \quad  W=W_1 \oplus W_2, 
\ee
and
\bel
&&A_i ={\bm B}_1\big|_{V_i} \in \text{End} (V_i), \quad  B_i={\bm B}_2 \big|_{V_i}  \in \text{Hom}(V_{i},V_{i+1}), \non
&&I_i={\bm I} \big|_{W_i} \in \text{Hom} (W_i, V_i), \quad J_i={\bm J}\big|_{V_i} \in \text{Hom} (V_i, W_{i+1})  .
\ee
The dimensions of $V_i, W_i$ are $\text{dim} \, V_i =k_i $, $\text{dim} \, W_1=N$, $\text{dim} \, W_2=N_f-N$.
Here the instanton number is $k_2$ and the vortex number is $k_1-k_2$. 
Then the moduli space is defined by a quotient:
\bel
\mathcal{M}_{\mathbb{L}, k_1, k_2}=\Bigl\{ (A_i, B_i, I_i, J_i) \Big| A_{i+1} B_i-B_{i} A_i + I_{i+1} J_i=0 \Bigr\} \Big/ \prod_{i} GL(k_i, \C) ,
\ee
with some stability condition. 
The equivariant character for the tangent space of $\mathcal{M}_{\mathbb{L}, k_1, k_2}$ is evaluated as
\bel
&&\chi (T_p \mathcal{M}_{\mathbb{L}, k_1, k_2})=\sum_{i,j=1}^2 \Bigl[ e^{\vep_1 +\frac{\vep_2}{2}} \chi (W_i) \chi (V^*_j) +\chi (V_i) \chi (W^*_j)
   \non
&&   ~~~~~~~~~~~~~~~~~~~~~~~~~~~~~~~~~~~~~~~~~~
 -(1-e^{\vep_1}) (1-e^{\frac{\vep_2}{2}})  \chi (V_i) \chi (V^{*}_j)  \Bigr] \Big|_{\Z_2-\text{even}},
\ee
with
\bel
&&\chi(W_1)=e^{\frac{\varepsilon_2}{2}} \sum_{a=1}^{N} e^{M_a}, \quad \chi(W_2)= \sum_{b=1}^{N_f-N} e^{{M'}_b} ,\non
&&\chi(V_1)=e^{\frac{\varepsilon_2}{2}} \sum_{a=1}^N e^{M_a} \sum_{ (i, 2j-1) \in Y_a} e^{-(i-1) \vep_1 - (j-1) \vep_2}
+\sum_{b=1}^{N_f-N} e^{{M'}_{b}} \sum_{ (i, 2j) \in X_b} e^{-(i-1) \vep_1 - (j-\frac{1}{2}) \vep_2} ,\non
&&\chi(V_2)=e^{\frac{\varepsilon_2}{2}} \sum_{a=1}^N e^{M_a} \sum_{ (i, 2j) \in Y_a} e^{-(i-1) \vep_1 - (j-1) \vep_2}
+\sum_{b=1}^{N_f-N} e^{{M'}_b} \sum_{ (i, 2j-1) \in X_b} e^{-(i-1) \vep_1 - (j-\frac{1}{2}) \vep_2} .\non
\ee
Here $Y_a$'s, $X_b$'s are Young diagrams.
The collection $(M_1 \cdots M_N, M^{'}_1,\cdots, M^{'}_{N_f-N})$ is  
the set of Coulomb branch parameters and 
 $\vep_i$'s are $\Omega$-background parameters.  
In the above equation for the equivariant character, 
the symbol ``$\Z_2\text{-even}$" means that we remove terms expressed as $e^{r\vep_2}$ $(r=\frac{1}{2},\frac{3}{2},\cdots)$.
The non-negative integers $k_i$'s are related to the  number of  boxes by the  following equations:
 \bel
k_1 =\sum_{a=1}^{N} \# \{ (i,2j-1)  | (i,2j-1) \in Y_a\}  
+ \sum_{b=1}^{N_f-N} \# \{ (i,2j)  | (i,2j) \in X_b\},  \\
k_2 =\sum_{a=1}^{N} \# \{ (i,2j)  | (i,2j) \in Y_a\}  
+ \sum_{b=1}^{N_f-N} \# \{ (i,2j-1)  | (i,2j-1) \in X_b\} ,
\label{instantonnumber} 
\ee
with $i,j \in \N$.
 Here $\# \{ \cdots \}$ expresses the cardinality of the set. 
 
Now we put the instanton number $k_2=0$. Then the ADHM data on $\C \times \C/\Z_2$  is reduced to $A_1$,$I_1$ and $J_1$:
\bel
&&A_1  \in \text{End} (V_1), \quad I_1\in \text{Hom} (W_1, V_1), \quad J_1 \in \text{Hom} (V_1, W_{2}).
\ee 
This  $(A_1,I_1, J_1)$ is precisely the same as the data of the moduli space of the 
$k_1$-vortex of the $G=U(N)$ gauge theory with the $SU(N_f)$ flavor symmetry \cite{Hanany:2003hp}. 
Moreover, from the relation (\ref{instantonnumber}), we find that  
$Y_a=\{ (i,1) | i=1,\cdots ,k^{a}_{1} \}$  with $\sum_{a=1}^N k^a_1=k_1$, 
but $X_b$'s are absent.
Thus the equivariant character $\chi (T_p \mathcal{M}_{\mathbb{L}, k_1, k_2})$ is reduced to
\bel
\chi (T_p \mathcal{M}_{\mathbb{L}, k_1, k_2=0})&&=  e^{\vep_1 +\frac{\vep_2}{2}} \chi (W_2) \chi (V^*_1) +\chi (V_1) \chi (W^*_1)
 -(1-e^{\vep_1})   \chi (V_1) \chi (V^{*}_1) \non
&&=\sum_{a,a^{'}=1}^N e^{M_a-M_{a'}} \sum_{i=1}^{k^a_1} e^{(k^{a^{'}}_1+1-i) \vep_1} + \sum_{b=1}^{N_f-N}  \sum_{a=1}^{N} e^{M^{'}_b-M_a} 
\sum_{i=1}^{k^a_1} e^{i \vep_1} ,
\label{chivortex}
\ee
with
\bel
&&\chi(W_1)=e^{\frac{\varepsilon_2}{2}} \sum_{a=1}^{N} e^{M_a}, \quad \chi(W_2)= \sum_{b=1}^{N_f-N} e^{{M'}_b}, \quad \chi(V_1)=e^{\frac{\varepsilon_2}{2}} \sum_{a=1}^N e^{M_a} \sum_{ i=1}^{k^{(a)}_1} e^{-(i-1) \vep_1 } .\non
\ee
$\chi (T_p \mathcal{M}_{\mathbb{L}, k_1, k_2=0})$ does not depend on the equivariant parameter $\vep_2$ and is precisely 
 the same  as the  equivariant character of the tangent space of the 
$k_1$-vortex moduli space of the $G=U(N)$ gauge theory with $N_f$-fundamental chiral multiplets. The latter has 
the fixed point $p$ labeled by $(k^1_1, \cdots, k^N_1)$ under the equivariant action.
In the context of  the three dimensional theory on $S^1 \times \C$, 
the Coulomb branch parameters correspond to real masses or fugacities associated to the $SU(N_f)$-flavor symmetry.  

The K-theoretic vortex partition function can be 
obtained from the equivariant character $ \sum_{i} \pm e^{ \omega_{i, p}}$
by  the replacement $\sum_{i } \pm e^{ \omega_{i,p}} \to \sum_{p} \prod_{i} (1-e^{\omega_{i, p}})^{\mp1}$ as 
in the case of the K-theoretic instanton partition  function.
Here $p=(k^1_1, \cdots, k^N_1)$ denotes a fixed point under the 
 $U(1)^{N_f-1}_{M} \times U (1)_{\vep_1} $- equivariant action  and  $\omega_{i,p}$'s denote the equivariant weights at the point $p$.  Then, from the equivariant character (\ref{chivortex}), 
the  vortex partition function with   vortex number $k_1$ is written as
\bel
Z_{k_1\text{-vortex}}=\sum_{\sum_a k^a_1=k_1}\prod_{a,a'=1}^N (1-e^{M_a-M_{a'}+(k^{a'}_1+1-i)\vep_1})^{-1} \prod_{b=1}^{N_f-N} \prod_{a=1}^N \prod_{i=1}^{k^a_1} (1-e^{{M'}_b-M_a +i \vep_1})^{-1}. \non
\label{vfunchi}
\ee
 When we identify the parameters as $e^{\vep_1}=q^{-2}$, $e^{{M'}_b}=z_{l}, (b+N=l)$ and $e^{M_{a}}=z_{l'}, (a=l')$, 
(\ref{vfunchi}) agrees with 
the vortex number $k_1$ sector of (\ref{ktheoreticvortex}) with $\tilde{N}_{f}=0$. 

\subsection{Calabi-Yau model and 3d matrix factorization}
  
So far all the models we considered do not have surface terms of superpotentials. In this section, we will consider a simple model
 which has non-trivial three dimensional analogue of the matrix factorization and study its two dimensional limit.

In this section we consider an analogue of an $\mathcal{N}=(2,2)$ GLSM which flows in the IR limit to
some non-linear sigma model whose target space is  a Calabi-Yau $(N-2)$-fold 
 $\text{CY}_{N-2}$ defined by a  degree $N$ homogeneous polynomial  $f(x_1,\cdots, x_N)=0$ in $\C \mathbf{P}^{N-1}$ 
in the large volume regime. 
 As a set-up, we consider an Abelian model with 
$G=U(1)$ and take a set of chiral multiplets $P, \phi_I,  (I=1,\cdots, N)$ with a superpotential
\bel
W(P, \phi_{I})=P \cdot f(\phi_I).
\ee
Here $f(\phi_I)$ is the homogeneous polynomial of a degree $N$.  
The charge assignments of the chiral multiplets are listed in Table \ref{CalabiYauN}.
We impose the Neumann boundary conditions for these chiral multiplets.
\begin{table}[htb]
\begin{center}
\begin{tabular}{|c||cc|}
\hline
  &   $U(1)_G$   & $U(1)_R$ \\
\hline
$P$  &   $-N$  & $+2$  \\
$\phi_I$  &   $+1$  & $0$  \\
 \hline
\end{tabular} 
\end{center}
\caption{The charge assignments of scalars in the three dimensional chiral multiplets.}
\label{CalabiYauN}
\end{table}
Then the one-loop contribution of the bulk three dimensional chiral multiplets
is given by
\bel
Z^{\text{3d.chi.N}}_{1\text{-loop}}=(e^{-i \beta r a}; q^2)^{-N}_{\infty} (e^{ i N \beta r a} q^{2}; q^2)^{-1}_{\infty}.
\ee
In addition, we have another contribution from the two dimensional boundary 
and the corresponding boundary theory is characterized by functions $E_a$ and $J^a$. 
This boundary effect is correlated to the bulk 3d theory through 
some kind of a three dimensional matrix factorization. 
We choose  $E_a$ and $J^a$ with $E_a J^a=W$ in order to realize this factorization
\bel
E(P, \phi_I)=P, \quad J(P, \phi_I)=f(\phi_I).
\ee
Then the boundary contribution in the partition function comes from
the one-loop determinant of the Fermi multiplet coupled to $E(P, \phi_I)$ and is given by
\bel
Z^{\text{2d.Fermi}}_{1\text{-loop}}=\th (e^{N i \beta r a} q^2; q^2).
\ee  
We can also include the FI-term and write down the expression of the partition function of the model
\bel
\mathcal{I}^{\text{CY}_{N-2}}_{S^1\times D^2}&&=\int \frac{d (\beta r a)}{2\pi } e^{-S_{FI}} Z^{\text{3d.chi.N}}_{1\text{-loop}} Z^{\text{2d.Fermi}}_{1\text{-loop}} \non
&&= \int \frac{d (\beta r a)}{2\pi } e^{2\pi  r \zeta ( i \beta r a)} \frac{\th (e^{N i \beta r a} q^2; q^2)}{(e^{-i \beta r a}; q^2)^{N}_{\infty} (e^{ i N \beta r a} q^{2}; q^2)_{\infty}} \nonumber \\
&&=\int \frac{d (\beta r a)}{2\pi } e^{2\pi  r \zeta ( i \beta r a)} \frac{ (e^{-N i \beta r a} ; q^2)_{\infty}}{(e^{-i \beta r a}; q^2)^{N}_{\infty} }.
\label{CYindex}
\ee
In the two dimensional limit $\beta\rightarrow 0$ 
with keeping $ t_{\text{2d}}:=2\pi r \beta \zeta $ finite,  (\ref{CYindex}) is reduced to 
\begin{align}
\lim_{\beta \to 0} ~ \mathcal{I}^{\text{CY}_{N-2}}_{S^1\times D^2} & \sim
\int \frac{d y}{2\pi i} e^{ t_{2d} \,y} \Gamma (1-N y)\Gamma(y)^{N} 
 ( e^{i N \pi y}-e^{-i N \pi y} ) \nonumber \\
&=
\int \!\! dy\, e^{ t_{2d} \,y} \, \frac{\Gamma(y)^{N}}{\Gamma (N y) }.
\label{eq:CYhemi}
\end{align}
Here we also defined $y:=i ra$.
The first line in \eqref{eq:CYhemi} agrees with the partition function on $D^2$ 
with a  loop operator wrapping on  $S^1=\partial D^2$ associated  with a matrix factorization, where the Neumann boundary condition is imposed for the chiral multiplets. 
The boundary condition of a one-loop determinant changes from the
Neumann-type to the Dirichlet-type
by the effect of the loop operator. Then we obtain  the second line in \eqref{eq:CYhemi} which is  related to the $\Gamma$-class of 
the Calabi-Yau $(N-2)$-fold $\text{CY}_{N-2}$.
Thus  \eqref{CYindex} is thought of as a 3d extension of 
the hemisphere partition function for 
$\text{CY}_{N-2}$ and the ratio of the q-Pochhammer symbols in \eqref{CYindex} can be regarded as 
a $q$-deformation of the $\Gamma$-class of $\text{CY}_{N-2}$.

\subsection{3d $\mathcal{N} \ge 3$ models and gauge/Bethe correspondence}
In this subsection, we study  three dimensional $\mathcal{N} \ge 3$ supersymmetric theories and Bethe ansatz for quantum integrable models.
We consider the $G=U(N)$ theory with the $N_f$ fundamental  hypermultiplets.
In the language of the $\mN=2$ multiplets, 
an $\mN=4$ vector multiplet consists of  an $\mN=2$   vector multiplet $(A_{\mu},\sig, D, \lam, \blam)$
and  an $\mN=2$  adjoint chiral multiplet  $(\sig', \lam', D')$. 
On the other hand, an $\mN=4$ fundamental hypermultiplet consists of  an $\mN=2$  fundamental chiral multiplet 
$(\phi, \psi, F)$ and an anti-fundamental chiral multiplet $(\tilde{\phi}, \tilde{\psi}, \tilde{F})$. 

\begin{table}[htb]
\begin{center}
\begin{tabular}{|c||ccccccc|}
\hline
 &   $A_{\mu}$  &   $\lam$ &  $\sig$  &$D$ & ${\sig '}$ & ${\lam '}$  & ${D '}$ \\
\hline
  $U(1)_R$  & $0$ & $1$  &   $0$ & $0$& $1$ &$0$  & $-1$ \\
   $U(1)_F$  & $0$ & $0$ &   $0$ & $0$& $1$  &$1$  &$1$  \\
\hline
\end{tabular} 
\end{center}
\caption{$U(1)_R \times U(1)_F$ R-charge assignments for the $\mN=4$ vector multiplet  }
\label{chargeN4vec}
\end{table} 

\begin{table}[htb]
\begin{center}
\begin{tabular}{|c||cccc|}
\hline
 &   $\phi$  &   $\psi$ &  ${\tilde{\phi}}$  &${\tilde{\psi}}$  \\
\hline
  $U(1)_R$  & $\frac{1}{2}$ & $-\frac{1}{2}$  &   $\frac{1}{2}$ & -$\frac{1}{2}$ \\
   $U(1)_F$  & $-\frac{1}{2}$ & $-\frac{1}{2}$ &   -$\frac{1}{2}$ & -$\frac{1}{2}$  \\
\hline
\end{tabular} 
\end{center}
\caption{$U(1)_R \times U(1)_F$ R-charge assignments for the $\mN=4$ hypermultiplet  }
\label{chargeN4hyper}
\end{table} 
The R-symmetry group of  the $\mathcal{N}=4$ supersymmetric theory is $SU(2)_L \times SU(2)_R$ and 
the   Cartan generator of $SU(2)_L (\text{resp. } SU(2)_R)$ is  $J_L (\text{resp. } J_R)$. 
The $U(1)_R$ R-symmetry in the $\mathcal{N}=2$ supersymmetry is generated by a generator $R$,  which is related 
to the $SU(2)_L \times SU(2)_R$ generators by $R=J_L-J_R$.
 On the other hand, the combination $F=J_L+J_R$ defines another global charge which commutes with the 
$U(1)_R$ and we introduce the fugacity $t$ for  $U(1)_F$.
The R-charge assignments for the $\mN=4$ multiplets are given in Table \ref{chargeN4vec} and Table \ref{chargeN4hyper}.
Now we impose the Neumann boundary conditions on the adjoint chiral multiplet $(\sig', \lam', D')$ and  on
the fundamental chiral multiplet $(\phi, \psi, F)$. We impose  
the Dirichlet boundary conditions\ on the anti-fundamental chiral multiplet
 $(\tilde{\phi}, \tilde{\psi}, \tilde{F})$. 
In this choice, all the anomalous terms from the three dimensional multiplets cancel each other, thus Chern-Simons levels are not shifted from the bare ones.
If we introduce the Chern-Simons term, the supersymmetry is broken to $\mathcal{N}=3$. 
When we take bare Chern-Simons levels as zero, the 3d index is well-defined without introducing any boundary multiplets.
Then the $\mathcal{N}=4$ index on $S^1 \times D^2$ is given by  
\bel
\mathcal{I}^{\mathcal{N}=4}_{S^1 \times D^2} &&=\frac{(t ;q^{2})^{-N}_{\infty}}{N ! } \int \prod_{a=1}^{N} \frac{d s_a}{2\pi is_a}   
 \prod_{1 \le a \neq b \le N}  \frac{ (s_a s^{-1}_{b} ;q^{2})_{\infty}} { (s_a s^{-1}_{b} t q; q^{2})_{\infty} }  
 \prod_{l=1}^{N_f}  \prod_{a=1}^N  \frac{ (s_a q^{\frac{3}{2} } t^{\frac{1}{2}} {z}_{l} ; q^2)_{\infty}}{( s_a t^{-\frac{1}{2}} q^{\frac{1}{2}} z_{l} ;q^2)_{\infty}} .
 \non
 \label{N4index}
\ee
This equation has an important property that the one-loop determinant of the 3d $\mN=4$ vector multiplet 
is expressed by the Macdonald measure and agrees with the index for the $\mathcal{N}=4$ theory on $S^1 \times \mathbb{C}$ computed in \cite{Aganagic:2013tta} by 
some  redefinitions of the fugacities.

Let us study the relation to integrable spin models. 
First we examine the relation between the $\mathcal{N}=4$ SQCD and the 
spin-$\frac{1}{2}$ XXZ quantum spin chain. 
It is shown in \cite{Nekrasov:2009ui} (see also \cite{Chen:2012we}) that  the saddle point equation of the twisted effective superpotential
 for the mass deformed $\mathcal{N}=4$ SQCD corresponds to the Bethe ansatz equation
 for the spin-$\frac{1}{2}$ XXZ model.  
We recall that the partition function on the three dimensional  ellipsoid $S^3_{b}$ depends on the squashing parameter $b$. In the limit $b \to 0$, 
an effective twisted superpotential appears.   
Similarly, we will see the effective twisted suprepotential appears in the  limit  $q \to 1$ with 
keeping the other parameters finite. This is different from the two dimensional limit.
In the Abelian case (the rank of the gauge group $N=1$), 
semiclassical behaviors of the holomorphic blocks  correctly reproduce the Bethe ansatz equation for the 
XXZ spin chain \cite{Gadde:2013wq}(see also \cite{Nieri:2013vba}). Then 
we expect the Bethe ansatz equation also appears  in the limit $q \to 1$   of the 
3d index  (\ref{N4index}) for the generic rank $N$. To see this, we take the limit $q \to 1$  and 
look at the behavior of the index:
\bel
\mathcal{I}^{\mathcal{N}=4}_{S^1 \times D^2} \sim \frac{1}{N!}\int \prod_{a=1}^{N} \frac{d s_a}{2 \pi i s_a }  ~ \exp{\Bigl( \frac{1}{2\beta} \mathcal{W}^{\mathcal{N}=4}_{\text{eff}} \Bigr) }, \quad (q \to 1),
\ee 
with
\bel
\mathcal{W}^{\mathcal{N}=4}_{\text{eff}}=\sum_{a \neq b} \left( \text{Li}_2 (s_a s^{-1}_b) -\text{Li}_2 (t s_a s^{-1}_b) \right)  
+ \sum_{l=1}^{N_f} \sum_{a=1}^N  \left( 
  \text{Li}_2 (s_a  t^{\frac{1}{2}} {z}_l) -\text{Li}_2 (s_a t^{-\frac{1}{2}} z_l) \right) .\non
\ee
The saddle point equation ``$\exp \left( s_a \partial_{s_a}  \mathcal
{W}^{\mathcal{N}=4}_{\text{eff}} \right) =1$'' is given
\bel
\prod_{b=1 \atop b \neq a}^N \frac{\sinh (y_b-y_a-c)}{\sinh(y_a-y_b-c)}
=\prod_{l=1}^{N_f} \frac{\sinh (y_a+\frac{c}{2} +M_l/2)}{\sinh (y_a-\frac{c}{2} +{M}_l/2)}.
\ee
Here we defined $2y_a=\log s_a$, $2c:=\log t$.
This is equivalent to the Bethe ansatz  equation 
for the $\mathfrak{sl} (2)$  spin-$\frac{1}{2}$ inhomogeneous XXZ quantum spin chain 
by some redefinitions of the  parameters.

The parameters in the gauge theory are related to those of the spin chain as follows; the rank of the gauge group $N$ 
corresponds to the number of excitations   and the number of the hypermultiplets $N_f$ corresponds to the number of sites of the spin chain. 
So far the  periodic boundary condition is imposed in the spin chain system. 
The twisted   boundary condition in the spin chain 
is  reproduced by introducing the FI-term. Also the generalization to the 
$\mathfrak{sl} (K)$ case is straightforward by considering a quiver gauge theory.  

The correspondence between  mass deformations of the $\mathcal{N}=4$ SQCD and the XXZ model is already known in the context of the gauge/Bethe correspondence. But we will propose 
a new example for the gauge/Bethe correspondence;
 the ``gauge" side is the pure $\mathcal{N}=3$ Chern-Simons-Matter theory, namely, 
we take $N_f=0$ in (\ref{N4index}) and introduce a dynamical Chern-Simons term 
with a level $\kappa$. The 
``Bethe" side (quantum integrable model) is the q-boson hopping model \cite{Bogoliubov:1993}.
The existence of the gauge/Bethe correspondence for Chern-Simons(-Matter) theories   was  pointed out first in \cite{Okuda:2012nx, Okuda:2013fea}. 
The boundary Chern-Simons terms are evaluated at the saddle point as
\bel
 e^{ S_{b.CS}} = \exp(-  \sum_{a=1}^{N} \frac{\kappa}{4\beta} (\log s_a)^2) .
\ee 
Thus we can read off the effective twisted superpotential 
\bel
\mathcal{W}^{\mathcal{N}=3}_{\text{eff}}=-\frac{\kappa}{2} \sum_{a=1}^{N} \log^2 (s_a)
+\sum_{a \neq b} \left( \text{Li}_2 (s_a s^{-1}_b) -\text{Li}_2 (t s_a s^{-1}_b) \right),
\ee
and obtain the set of saddle point equations
\bel
s_a^{\kappa}=\prod_{b=1 \atop b \neq a}^N \frac{s_a t-s_b}{s_a- t s_b}.
\ee 
This reproduces the Bethe ansatz equation for the $q$-boson hopping model in the $N$-particles sector with the periodic boundary condition. The number of sites  corresponds to the Chern-Simons level $\kappa$.

\subsection{Domain wall index on $S^1 \times S^2$}
In four dimensions, 4d-3d coupled partition functions or superconformal indices have been introduced in \cite{Drukker:2010jp}, \cite{Gang:2012ff}.   
In this section we briefly mention on a 3d-2d domain wall index on $S^1 \times S^2$.

The $\mathcal{N}=(0,2) $ domain wall theory lives on $T^2= S^1 \times S^1_{\mth=\frac{\pi}{2}} \subset S^1 \times S^2$ and 
 couples to two different theories, each of which lives on a  northern part $S^1 \times D^2_N (\mth \le \frac{\pi}{2} )$ 
or a southern part  $S^1 \times D^2_S ({\mth \ge \frac{\pi}{2}} )$. 
Let $G_{N}$ and $G_{S}$   be gauge groups of the $\mathcal{N}=2$ theories on   $S^1 \times D^2_{N}$ 
and  $S^1 \times D^2_{S}$, respectively.  
Let  $Z_{S^1\times D^2_{N}}^{1\text{-loop}} (s)$, $Z_{S^1\times D^2_{S }}^{1\text{-loop}} (\tilde{s})$
 be the one-loop determinants of the $\mathcal{N}=2$ theories on the northern   $S^1 \times D^2_{N}$ and the 
 southern $S^1 \times D^2_{S}$.
 Here  $s_a$'s and  $ \tilde{s}_b$'s label Cartan parts of exponentiated holonomies of $G_N$ and $G_S$ along the 
$S^1$-direction.  On the boundary torus,  
the multiplets of the $\mathcal{N}=(0,2) $ theory have charges associated with  the gauge groups $G_N \times G_S$ 
and its partition function $Z_{T^2}^{1\text{-loop}} (s, \tilde{s})$ is given by 
the one-loop determinant of the  $\mathcal{N}=(0,2)$ theory.
By collecting these functions, we can write down the domain wall index: 
\bel
\mathcal{I}^{\text{DW}}_{S^1 \times S^2} = \int \int \prod_{a=1}^{\text{rk}(G_N)} \frac{d {s_a}}{2\pi s_a} 
\prod_{b=1}^{\text{rk}(G_S)} \frac{d {\tilde{s}_b}}{2\pi \tilde{s}_b} 
Z_{S^1\times D^2_{N }}^{1\text{-loop}} (s) Z_{T^2}^{1\text{-loop}} (s, \tilde{s}) 
Z_{S^1\times D^2_S}^{1\text{-loop}} (\tilde{s}). 
\ee 
The matter contents and two bare Chern-Simons levels have to be chosen to cancel  the  total bulk-boundary anomalies.
 
\subsection{Wilson-Vortex loop and $q$-shift operator}
In this subsection we study properties of flavor Wilson loops and vortex loops on the 
3d index  $\mathcal{I}_{S^1 \times D^2} (q^2,z)$.  
When we turn on the $l$-th flavor background gauge field ${A}^{(l)}_{\mu}=(0, 0, -i M_{l}/\beta r)$, 
an associated  flavor Wilson loop with a charge $Q$ is defined by
\bel
W^{(l)}_F =\exp \left( i Q \int {A}^{(l)}_{\tau} d \tau  \right).
\ee
Then the expectation value of the flavor Wilson loop is given by using the index $\langle W^{(l)}_F  \rangle =z^{-Q}_l \mathcal{I}_{S^1 \times D^2}$ with $z_l=e^{M_l}$. 
The localization computation of  vortex loops  was  studied in \cite{Kapustin:2012iw, Drukker:2012sr} for  $S^3$ and $S^1 \times S^2$. 
The vortex loops are defined as some defect operators specified by  line singularities similar to the 't Hooft loops.
This means the boundary conditions for the component fields in three dimensional theories are modified near the vortex loops.
Since the appropriate equivariant index theorem for the manifold with the boundary has not been known yet, 
it is  difficult to  directly evaluate the effects of the vortex loops on the manifold with the boundary, for example, 
$S^1 \times D^2$.
Instead of the direct computation, we  apply the method studied in  \cite{Kapustin:2012iw} to our model. 
That is, the vortex loop is obtained by acting an $S$-transformation  on the flavor Wilson loop:
\bel
V^{(l)}_F =S^{-1} W^{(l)}_F S, \quad  S \in SL(2,\Z).
\ee 
Here the $S$-transformation is defined by adding an 
FI-term for the $l$-th flavor background gauge field and by gauging this background field 

 \bel
( S \cdot \mathcal{I}_{S^1 \times D^2} ) (q^2, \zeta_{l}) = \int ~ d M_l ~ \mathcal{I}_{S^1 \times D^2}  ~ e^{2\pi i \zeta_l M_l} .
\ee
The transformation $S^{-1}$ is also given as the inverse transformation of $S$
and  the vortex loop acts on the 3d index as
\bel
(S^{-1} W^{(l)}_F S ) \cdot \mathcal{I}_{S^1 \times D^2} (q, z^{'}_l) && =\int d \zeta_l  e^{-2 \pi i M^{'}_{l} \zeta_l}  e^{2\pi i Q \zeta_l} \int ~
d M_l ~ \mathcal{I}_{S^1 \times D^2}  ~ e^{2\pi i \zeta_l M_l} \non
&&=\mathcal{I}_{S^1 \times D^2} (q^2, z^{'}_{l} q^Q).
\ee 
 Thus we obtain the 
 expectation value of the vortex loop for the $l$-th flavor gauge field 
by shifting the $l$-th flavor fugacity $\langle V^{(l)}_F  \rangle 
= \mathcal{I}_{S^1 \times D^2} (q^2,z_l q^Q)$. As a result, the flavor vortex loops are regarded as $q$-shift operators.
 The successive actions of Wilson-vortex loops on the 3d index do not commute, but they satisfy
the following  commutation relations
 \bel
 \langle  V^{(l)}_F  W^{(k)}_F \rangle=q^{-Q} \delta_{l k} \langle  W^{(k)}_F  V^{(l)}_F \rangle .
\ee

\section{Summary and discussion }
In this paper, we have evaluated the partition functions of the $\mN=2$ supersymmetric Chern-Simons-Matter theories on 
$S^1 \times D^2$ in terms of localization techniques. In the particular choice of the fugacity $\beta_2 = \beta $,  we find  
the conditions to cancel anomalous terms are reduced to the decomposition rule 
for effective mixed Chern-Simons levels in the holomorphic blocks for  Abelian gauge theories. 
In these cases, our 3d-2d indices  reproduce the holomorphic blocks.  
On the other hand, in the non-Abelian gauge theories,
there might be a mismatch in the sector of the vector multiplet and 
 effective Chern-Simons levels.
 One possibility of the mismatch comes from the difference in the metrics of 
$S^1 \times D^2$ and Melvin cigar. Both spaces have the same topology as the solid torus, but have different metrics. 
It is desirable to study this point further to reveal  the origin of this discrepancy. 
We postpone this problem in our future work. 

We have also studied the connection between our indices on $S^1 \times \D$ and 
several topics; the K-theoretic $J$-function for the $\C \mathbf{P}^N$ model, 
vortex partition functions and surface operators, 
the 3d analogue of the matrix factorization, the gauge/Bethe correspondence and loop operators.

We  have constructed boundary interactions which can be regarded as three dimensional analogue of the matrix factorization. 
Although  the boundary interactions in three dimensions (the $\mathcal{N}=(0,2)$ superpotential term (\ref{fsuperpot})) have  quite different expressions from 
 two dimensional  ones (roughly speaking, Wilson loops for superconnections), after the localization computation is performed, 
the partition functions can reproduce the two dimensional partition functions on the hemisphere by the dimensional reduction.

In this paper we did not study the boundary interaction described by the 
$G/G$ chiral gauged WZW model in detail. It is interesting to study the 
3d-2d index  with this boundary interactions. 

In \cite{Gadde:2013sca}, the 2d-4d correspondence was proposed.
The 2d side describes the $\mathcal{N}=(0,2)$  flavored elliptic genus \cite{Benini:2013nda, Gadde:2013dda} 
and the 4d side is related to the  Vafa-Witten partition functions \cite{Vafa:1994tf} on four-manifolds. 
In this paper, we have realized  $\mathcal{N}=(0,2)$  theories as the boundary interactions of 3d $\mathcal{N}=2$ supersymmetric theories.  
The 3d-2d coupled index is expected to be related to some Vafa-Witten 
partition function with degrees of freedom on the three dimensional boundary, 
which is  realized as the 
asymptotic boundary  of the four-manifold. 
It is interesting to explore the connection between the 3d-2d indices and the partition functions of the 3d-4d coupled systems.

\subsection*{Acknowledgment}
YY is grateful to Richard~Eager,  Dongmin Gang,  
 Kentaro~Hori, Bumsig~Kim, Kimyeong~Lee, Yu~Nakayama, Mauricio~Romo, Yuji~Tachikawa, Masato~Taki, Akinori~Tanaka, Seiji~Terashima, Kazushi~Ueda, Futoshi~Yagi  
and  Satoshi~Yamaguchi  for useful comments or discussions. He also thanks Richard~Eager and Kentaro~Hori for kind hospitality during visiting Kavli IPMU.

\appendix
\section{Conventions of 3d $\mN=2$ theory on $S^1 \times \D$}
We use gamma matrices $\gamma_{\hat{a}}$ in the local Lorentz frame:
\bel
\gamma_{\hat{1}} = \begin{pmatrix} 0 & 1 \\ 1 & 0  \end{pmatrix}, \quad 
\gamma_{\hat{2}} = \begin{pmatrix} 0 & -i \cr i & 0\cr \end{pmatrix}, \quad
\gamma_{\hat{3}} = \begin{pmatrix} 1 & 0 \cr 0 & -1\cr \end{pmatrix}.
\ee
In curved spaces, one can define 
 $\gamma_{\mu}=e^{\hat{a}}_{\mu} \gamma_{\hat{a}}$ by using 
a dreibein $e^{\hat{a}}_{\mu}$.
The charge conjugation matrix is expressed by
\bel
C_{\alpha \beta} = -i\gamma_{\hat{2}} = 
\begin{pmatrix} 0 & -1 \cr 1 & 0\cr \end{pmatrix},
\ee
which satisfies $C \gamma^{\mu} C^{-1}=-(\gamma^{\mu})^T$.
With this $C$, the spinor product is defined by
\bel
\ep \psi:= \ep^{\alpha} \psi_{\alpha} = \ep^{\alpha} C_{\alpha \beta} \psi^{\beta}=\ep^{T} C \psi.   
\ee

The supersymmetric transformation of the $\mathcal{N}=2$  vector multiplet  is given by
\bel
&&\delta A_{\mu}=\frac{i}{2} (\bar{\ep} \gamma_{\mu} \lambda - \bar{\lambda} \gamma_{\mu} \ep), \non
&&\delta \sigma=\frac{1}{2} (\bar{\ep}  \lambda - \bar{\lambda}  \ep), \non
&&\delta \lambda=-\frac{1}{2} \gamma^{\mu \nu} F_{\mu \nu} \ep -D \ep+i \gamma^{\mu}D_{\mu} \sigma \ep+\frac{2i}{3} \sigma \gamma^{\mu} D_{\mu} \ep ,\label{SUSYtransvec1} \\
&&\delta \bar{\lambda}=-\frac{1}{2} \gamma^{\mu \nu} F_{\mu \nu} \bep +D \bep-i \gamma^{\mu}D_{\mu} \sigma \bep
-\frac{2i}{3} \sigma \gamma^{\mu} D_{\mu} \bep ,\non
&&\delta D=-\frac{i}{2} \bep \gamma^{\mu} D_{\mu} \lambda-\frac{i}{2} D_{\mu} \bar{\lambda} \gamma^{\mu}  \ep
+\frac{i}{2}[\bep \lambda, \sigma]+\frac{i}{2}[\bar{\lambda} \ep , \sigma]
-\frac{i}{6} (D_{\mu} \bep \gamma^{\mu} \lambda + \bar{\lambda}\gamma^{\mu} D_{\mu} \ep  ) .
\nonumber
\ee 
For the $\mathcal{N}=2$  chiral multiplet, the supersymmetric transformation is expressed as
\bel
&&\delta \phi= \bep \psi ,\non
&&\delta \bphi= \ep \bpsi ,\non
&&\delta \psi= i \gamma^{\mu} \ep D_{\mu} \phi+i \ep \sig \phi+\frac{2i \Delta}{3} \gamma^{\mu} D_{\mu} \ep \phi+\bep F ,\non
&&\delta \bpsi= i \gamma^{\mu} \bep D_{\mu} \bphi+i \bphi \sig \bep+ \frac{2i \Delta}{3} \bphi \gamma^{\mu} D_{\mu} 
 \bep +\bF \ep , \\
&&\delta F=\ep (i \gamma^{\mu} D_{\mu} \psi-i \sig \psi -i\lam \phi)+\frac{i}{3} (2\Delta-1)  D_{\mu} \ep \gamma^{\mu} \psi, \non
&&\delta \bF=\bep (i \gamma^{\mu} D_{\mu} \bpsi-i \bpsi \sig  +i \bphi \bar{\lam} )+\frac{i}{3} (2\Delta-1)  D_{\mu} \bep \gamma^{\mu} \bpsi .\nonumber 
\ee
Here the covariant derivative is defined by $D_{\mu}=\nabla_{\mu}+i A^{a}_{\mu} T^a_{\mathcal{R}}$ with the vector boson $A_{\mu}=A^a_{\mu}T^a_{\cal R}$ 
in the representation ${\cal R}$.

\section{2d $\mathcal{N}=(0,2)$ supersymmetry on the boundary torus}
The generators of the supersymmetric transformations are defined 
by the restriction of  the three dimensional Killing spinors $\epsilon'$ and $\bar{\epsilon}'$ on the boundary torus  :
\bel
&&(D_{\hat{2}}+i D_{\hat{3}}) \ep'=\frac{i}{r} \ep', \quad (D_{\hat{2}}-i D_{\hat{3}}) \ep'=0, \non 
&&(D_{\hat{2}}+i D_{\hat{3}}) \bep'=\frac{-i}{r} \bep', \quad (D_{\hat{2}}-i D_{\hat{3}}) \bep'=0.
\ee
The set of commutators of the supersymmetric transformations of the vector multiplet 
is given by
\bel
&&[\delta_{1}, \delta_2] (A_{\hat{2}}-iA_{\hat{3}})=\alpha (-2i){F_{\hat{2} \hat{3}}}, \non
&&[\delta_{1}, \delta_2] \lam_1=\alpha \left[-2 (D_{\hat{2}}+i D_{\hat{3}}) \lam_1+\frac{2i}{r} \lam_1\right], \non
&&[\delta_{1}, \delta_2] \blam_1=\alpha \left[-2 (D_{\hat{2}}+i D_{\hat{3}}) \blam_1-\frac{2i}{r} \blam_1\right], \non
&&[\delta_{1}, \delta_2] \hat{D} =\alpha \left[2 (D_{\hat{2}}+i D_{\hat{3}}) \hat{D}\right].
\ee
Here we defined $\alpha={\bep '}_2 {\ep '}_1-{\bep '}_1 {\ep '}_2$.
Next the set of commutators of the supersymmetric transformations of the chiral multiplet 
is expressed as
\bel
&&[\delta_{1}, \delta_2] \phi= \alpha \left[2 (D_{\hat{2}}+i D_{\hat{3}}) \phi + 2 \frac{i\Delta}{r} \phi \right], \non
&&[\delta_{1}, \delta_2] \psi'= \alpha \left[-2 (D_{\hat{2}}+i D_{\hat{3}}) \psi' -\frac{2 i}{r} (\Delta-1) \psi' \right], \non
&&[\delta_{1}, \delta_2] \bphi= \alpha \left[ 2 (D_{\hat{2}}+i D_{\hat{3}}) \bphi - 2 \frac{i\Delta}{r} \bphi \right], \non
&&[\delta_{1}, \delta_2] \bpsi'= \alpha \left[-2 (D_{\hat{2}}+i D_{\hat{3}}) \bpsi' + \frac{2 i}{r} (\Delta-1) \bpsi' \right].
\ee
The set of commutators of the supersymmetric transformations of the Fermi multiplet 
is given by
\bel
&&[\delta_{1}, \delta_2] \Psi =\alpha \left[2 (D_{\hat{2}}+i D_{\hat{3}}) \Psi+\frac{2i}{r} (\tilde{\Delta}-1) \Psi \right], \non
&&[\delta_{1}, \delta_2] G =\alpha \left[2 (D_{\hat{2}}+i D_{\hat{3}}) G+\frac{2i}{r} (\tilde{\Delta}-2) G \right], \non
&&[\delta_{1}, \delta_2] \bar{\Psi} =\alpha \left[2 (D_{\hat{2}}+i D_{\hat{3}}) \bar{\Psi}+\frac{2i}{r} (1-\tilde{\Delta}) \bar{\Psi} \right], \non
&&[\delta_{1}, \delta_2] \bar{G} =\alpha \left[2 (D_{\hat{2}}+i D_{\hat{3}}) \bar{G}+\frac{2i}{r} (2-\tilde{\Delta}) \bar{G} \right].
\ee
\section{Definitions of  functions}

The dilogarithm function is defined by
\bel
\text{Li}_2 (x)=\sum_{n=1}^{\infty} \frac{x^n}{n^2},
\ee
and an integral representation of this dilogarithm function is given by
\bel
\text{Li}_2 (x)=-\int^{x}_{0} dt ~  \frac{\log (1-t)}{t}.
\ee
Next the q-Pochhammer symbol is defined by
\bel
(a;q)_{n}=\prod_{i=0}^{n-1} (1-aq^{i}),
\ee
and the q-theta function is defined for $|q|<1$
\bel
\th(y;q)=\prod_{n=0}^{\infty} (1-y q^{n} ) (1-y^{-1} q^{n+1}), \quad y \in \mathbb{C}^*, \, |q| <1.
\ee
The quantum dilogarithm function is defined by
\bel
\mathrm{Li}_2(x;q)=\sum_{n=1}^{\infty} \frac{x^n}{n(1-q^n)}, \quad  |x|, |q|  <1 .
\ee
The q-Pochhammer symbol is expressed by the quantum dilogarithm as 
\bel
(x;q)_{\infty} = \exp (-\mathrm{Li}_2(x;q)),
\ee
and the semiclassical limit is given by
\bel
\mathrm{Li}_2(x; e^{2\hbar}) \sim - \frac{1}{2 \hbar} \mathrm{Li}_2(x), \quad \hbar \to 0. 
\ee

\section{Derivation of one-loop determinants}
\label{AppendixD}
\subsection{3d vector multiplet}
In this subsection, we  evaluate one-loop determinants of super Yang-Mills theory (\ref{SYMaction}).
The evaluation of the one-loop determinant on $S^1 \times D^2$ can be performed in the similar manner to  \cite{Honda:2013uca, Hori:2013ika}. 
Because we treat bosonic fields on $S^1\times D^2$, we  introduce 
 scalar harmonics  $Y_{jm}$ and vector harmonics $(C^{\lambda}_{j m})_{i}$ on $S^2$ 
labeled by sets of $(j,m)$'s with $j\leq |m|$ $(j=1,2,3,\cdots)$.
Each field in the multiplet is expanded by these  harmonics
and generators $E_{\alpha}$'s  associated with roots $\alpha$'s in the Cartan Weyl basis: 
(Cartan parts for the fluctuations are omitted)

\bel
&&\sigma=\sum_{j}  \mathop{{\sum}'}_m \sum_{\alpha >0} \sig^{\alpha} Y_{jm}  E_{\alpha}+\text{(h.c)}, \\
&&A_i= \sum_{\lambda=1}^{2} \sum_{j=1}^{\infty} \mathop{{\sum}'}_{m=-j}^{j} \sum_{\alpha > 0} A^{ \alpha \lambda}_{jm} (C^{\lambda}_{j m})_{i} E_{\alpha} 
+\text{(h.c)}, \quad (i=1,2) \\
&&A_3=\sum_{\alpha > 0} \sum_{j=1}^{\infty} \sum_{m=-j \atop j-m=\text{even}}^{j} A^{ \alpha 3}_{jm} Y_{jm} E_{\alpha}+\text{(h.c.)}.
\ee
Here the symbol ``$(\text{h.c})$" denotes the Hermitian conjugate and 
the sum $\mathop{{\sum}'}$ runs over the
following modes under the boundary condition (\ref{veccondition}):
\bel
&&Y_{jm}: j-m=\text{even}, 
\quad (\text{Neumann}), \\
&&Y_{jm}: j-m=\text{odd},  
\quad (\text{Dirichlet}), \\
&& C^{1}_{j m }: j-m=\text{even},  
\quad C^{2}_{j m }: j-m=\text{odd}.  
\ee

Next we turn to the ghost fields $(c,\bar{c})$. 
The theory has gauge symmetry and we need to introduce ghost terms for fixing the symmetry
\bel
\mathcal{L}_{\text{ghost}+\text{g.f.}}=-\bar{c} \nabla^{i} D_{i} c 
+\frac{1}{2 \xi} (\nabla^{i} A_{i})^2.
\ee
We take the Neumann boundary condition for the ghost fields $(c, \bar{c})$ and expand them by the harmonics
\bel
&&c=\sum_{j}  \mathop{{\sum}'}_m\sum_{\alpha \neq 0} c^{\alpha}_{j m} Y_{j m} E_{\alpha}.
\ee
One can express the ghost part ${\cal L}_{\text{ghost}}$ and 
the bosonic part ${\cal L}_{\text{vec.b}}$ of super Yang-Mills  at the quadratic order as
\bel
&&\int_{D^2}{\cal L}_{\text{ghost}}=\frac{1}{2r^2}\sum_{\alpha}\mathop{{\sum}'}_{j,m}  \mbox{Tr}(E_{\alpha}E_{-\alpha})\cdot (-1)^m\,
j(j+1)\,\bar{c}^{\alpha}_{j,m}c^{-\alpha}_{j,-m},\non
&&\int_{D^2}{\cal L}_{\text{vec.b}}=\frac{1}{2r^2}\sum_{\alpha >0}\mathop{{\sum}'}_{j,m}\mbox{Tr}(E_{-\alpha}E_{\alpha})\cdot 
{\cal V}^{\alpha\dagger}_{jm}\cdot {\cal M}\cdot {\cal V}^{\alpha}_{jm}\,,
\ee
with
\bel
&&{\cal V}^{\alpha}_{jm}=(
\begin{array}{cccc}
A^{\alpha 1}_{jm} & A^{\alpha 2}_{jm} & A^{\alpha 3}_{jm} & \sigma^{\alpha }_{jm} 
\end{array})^T\,,\nonumber
\ee
\begin{eqnarray}
{\cal M}=
\begin{pmatrix}
 \xi^{-1}\cdot j(j+1) -r^2 D^3 D_3     &      0                   &     \sqrt{j(j+1)}\cdot rD_3                   & 0  \cr
   0                       & j(j+1) -r^2D^3 D_3     &   0                      & -\sqrt{j (j+1)} \cr           
   -\sqrt{j(j+1)}\cdot rD_3        &        0                 &  j(j+1)    & 0  \cr
   0                      &       -\sqrt{j (j+1)}    &       0                  &   j(j+1)+1 -r^2D^{3} D_{3}   \cr
\end{pmatrix}
. \nonumber \\
\end{eqnarray}

Then one-loop determinants of the bosonic and the ghost parts respectively become products of the modes $(j,m)$
\bel
&&Z_{\text{vec}.\text{b}}=\prod_{\alpha \neq 0} \prod_{j-m=\text{even}} \left[ j(j+1) \right]^{-1} \non
 && \qquad \times \prod_{\alpha \neq 0} \prod_{j-m=\text{odd}}
[ (j+1+r D_3) (j-rD_3) (j+1-rD_3) (j+rD_3) ]^{-\frac{1}{2}} ,
\label{1loopvecboson}\\
&&Z_{\text{ghost}}=\prod_{\alpha \neq 0} \prod_{j-m=\text{even}} \left[ j(j+1)\right].
\ee
Here $Z_{\text{ghost}}$ is cancelled by the first line  in the right hand side of (\ref{1loopvecboson}).

Next we shall evaluate eigenvalues of the operator $D_3$.
Let ${\cal O}_{n,m}$  be an operator  (field) with an R-charge $R$ and a flavor charge $F_l$, which  satisfies 
\bel
 \partial_{\tau} {\cal O}_{n,m}&=\frac{2 \pi i n}{\beta r} {\cal O}_{n,m},  \quad
 {\sf J}_3 {\cal O}_{n,m}= m {\cal O}_{n,m}.
\ee 
Then $D_3$ acts on ${\cal O}_{n,m}$ as  
\bel
\beta r D_3{\cal O}_{n,m}= \left[ 2 \pi i n + i \beta r \rho(a) -(R+m) \beta_1 +m \beta_2 +F_l M_l \right]{\cal O}_{n,m} \label{b-twist}.
\ee 
For example, the $R$-charge for the gauge field is $0$ and that for the fermion $\lambda$ (resp. $\blam$) is $-1$ (resp. $ +1$).
Then the  one-loop determinant of the bosonic part is given by
\bel
\prod_{\alpha \neq 0} \prod_{n \in \Z} \prod_{j=1}^{\infty} \prod_{\stackrel{\scriptstyle m=-j+1}{j-m=odd}}^{j-1} &&
(2 \pi i n+ i \beta r\alpha(a) +(j-m+1)\beta_1+(j+m+1)\beta_2 )^{-1} \non
&&  \times
(2 \pi i n+ i \beta r\alpha(a) +(j+m)\beta_1+(j-m)\beta_2 )^{-1}. \nonumber
\ee 
We make a remark here;
we have fixed the gauge symmetry with $\nabla^iA_i=0$, but there is a residual gauge symmetry $\delta A_{\tau}=D_{\tau}\kappa$
with a parameter $\kappa (\tau)$. In order to fix this symmetry, we impose a condition $\del_{\tau}\omega =0$ with 
$\omega :=\frac{1}{\text{vol}(D^2)}\int_{D^2}A_{\tau}$
and introduce a set of a ghost and an anti-ghost. 
When we integrate these ghost fields, they induce 
a contribution $\det D_{\tau}$ to the partition function. 
It is evaluated up to an overall  constant: 
\bel
&&\det D_{\tau}=\prod_{n\neq 0}\prod_{\alpha}\left(2\pi i\frac{n}{\beta r}+i\alpha (\lambda)\right)
\approx \prod_{\alpha>0}\frac{1}{\alpha (\lambda)^2}\sin^2\frac{\beta r\alpha (\lambda)}{2} ,
\ee
where $\lambda_i$ $(i=1,2,\cdots ,N)$ is the set of eigenvalues of the matrix $\omega$ and $\alpha$ is the root.
The measure of the matrix integral is expressed by  
$d\omega =
\prod_{i}d\lambda_i \prod_{\alpha >0}
\alpha (\lambda)^2$ and $d\omega \cdot \det D_{\tau}=
\prod_id\lambda_i \prod_{\alpha >0}\sin^2 \frac{\beta r \alpha (\lambda)}{2}$.

Next we shall consider the contribution of the fermions to the one-loop determinant. 
We consider the fermionic part of the Yang-Mills Lagrangian and evaluate the fluctuations around the saddle point 
at the quadratic order.
We expand the gaugino in terms of the spinor harmonics $\chi^{\pm}_{j,m} (\mth, \vphi)$:
\bel
&&\lam=\sum_{\alpha \neq 0} \sum_{s=\pm} \sum_{j}\mathop{{\sum}'}_{m}  \lam^{ \alpha, s}_{j m} \chi^{s}_{j m} E_{\alpha}. 
\ee
Here the sum $\mathop{{\sum}'}$ runs over the following modes under the boundary condition (\ref{veccondition}):
\bel
&& r\gamma_3 \gamma^{i} D_{i} \chi^{\pm}_{j m}=\pm \left(j+\frac{1}{2} \right) \chi^{\pm}_{j m} ,\non
&& \chi^{+}_{j m}: j-m=\text{even}, \quad
  \chi^{-}_{j m}: j-m=\text{odd}.
\ee
By using this spinor harmonics, 
we can write down 
the fermion part of the Yang-Mills Lagrangian
\bel
S^{(2)}_{\text{vec}} \Big|_{\text{fer}} 
 &&=\int_{S^1}  \sum_{\alpha \neq 0} \sum_{j=\frac{1}{2}}^{\infty}   \frac{i}{4r} 
\Bigl[ 
\sum_{m:j-m=\text{even}}   (-1)^{-m+\frac{1}{2}} \blam^{\alpha, -}_{j, -m}  ( j +     r D_{3}  )\lam^{-\alpha, +}_{j, m} \non
 &&
\qquad \qquad \qquad  +\sum_{m:j-m=\text{odd}} (-1)^{-m+\frac{1}{2}} \blam^{\alpha, +}_{j, -m}  ( j -    r D_{3} +1  )\lam^{-\alpha, -}_{j, m}
 \Bigr] \mbox{Tr}(E_{\alpha}E_{-\alpha}),\nonumber
\ee 
and calculate the one-loop determinant as
\bel
&& \prod_{j=\frac{1}{2}}^{\infty}  \prod_{m:j-m=\text{even}} \text{Det} ( j   +  r D_{3} )
\prod_{m:j-m=\text{odd}} \text{Det}  ( j -   r D_{3}  +1) 
 \label{vecfermode}.
\ee
First we can evaluate
the factor $\text{Det} (j  -  r D_{3}+1)$ with $j=j'+\frac{1}{2}$, $m=m'+\frac{1}{2}$:
\bel
\text{Det} ( \beta(j+1)-    \beta r D_{3} )
&&=\prod_{n \in \Z} \prod_{\alpha\neq 0} \prod_{j'=0}^{\infty} \prod_{\stackrel{\scriptstyle m'=-j'-1}{j'-m'=\text{odd}}}^{j'-1}
(2\pi i n+i \beta r \alpha (a)+(j'+m'+1)\beta_1 +(j'+1-m')\beta_2) .\nonumber
\ee
Similarly the other factor $\text{Det}( j  +  r D_{3} )$  
is evaluated with $j=j'-\frac{1}{2}$, $m=m'+\frac{1}{2}$:
\bel
\text{Det}( \beta j  + \beta r D_{3} )
&&=\prod_{n \in \Z} \prod_{\alpha\neq 0} \prod_{j'=1}^{\infty} \prod_{\stackrel{\scriptstyle m'=-j'+1}{j'-m'=\text{odd}}}^{j'-1}
(2\pi i n+i \beta r \alpha (a)+(j'-m')\beta_1 +(j'+m')\beta_2 ) .\nonumber 
\ee
Then the   one-loop determinant of the vector multiplet results in  the product formula
\bel
Z^{\text{3d.vec}}_{1\text{-loop}} 
&&=\prod_{\alpha\neq 0} e^{\frac{-(i \beta r \alpha (a))^2}{8\beta_2}} (q^2 e^{- i\beta r \alpha(a)};q^2).
\ee 
Here we adopted the zeta function regularization used in \cite{Tanaka:2014oda}. 
In the evaluation of the one-loop determinant in the following subsections, 
we use the common regularization scheme.
As expected, the one-loop determinant of the vector multiplet does not depend on the fugacity $\beta_1$.

\subsection{3d chiral multiplet}
\subsubsection{Neumann boundary condition}
We first evaluate the  one-loop bosonic determinant for the chiral multiplet. When the 
Neumann boundary condition (\ref{Ncondition}) is imposed, 
 $\phi$ can be expanded as follows: 
\bel
&&\phi=\sum_{\rho } \sum_{j=0}^{\infty} \sum_{m=-j\atop j-m=\text{even}}^{j} \phi^{\rho}_{ j m}  Y_{j m} (\mth, \vphi) E_{\rho}.   
\ee 
Here $\rho $ runs over the weight of the representation $\mathcal{R}$ of the Lie algebra $\mathrm{Lie} (G)$.  
At the quadratic order of fluctuations, the action of the chiral multiplet is expanded in terms of the scalar harmonics
\bel
&&{S}^{(2)}_{\text{chi}} \Big|_{\text{bos}} 
=\frac{1}{2 r^2} \int_{S^1}
\sum_{j=0}^{\infty} \sum_{m=-j \atop j-m=\text{even}}^{j} \sum_{\rho} \bar{\phi}^{\rho}_{ j, m}  
\left(   j +\Delta +  r D_{3} \right) \left(  j+1 -\Delta -  r D_3 \right)   \phi^{\rho}_{j m}.  \non
\label{chiralNmode}
\ee
Under the twisted boundary condition (\ref{twistedPB}), 
the factor $ \left(   j +\Delta+r D_{3} \right)$ in (\ref{chiralNmode}) contributes to 
the  one-loop determinant of the bosonic fields as  
\bel
&&\prod_{j=0}^{\infty} \prod_{m=-j \atop j-m=\text{even}}^{j}  
\mathrm{Det} \left(   (j +\Delta) \beta+ \beta r D_{3} \right){}^{-1}  \non
&&=\prod_{n \in \Z} \prod_{j=0}^{\infty} \prod_{m=-j \atop j-m=\text{even}}^{j} \prod_{l} 
(2\pi i n+i \beta r \rho(a)+(j-m) \beta_1+(j+\Delta+m) \beta_2 +F_l M_l){}^{-1} .\non
\ee
Similarly the other factor $\left(   j+1 -\Delta - r D_3 \right)$ contributes to 
the  one-loop determinant of the bosonic fields as 
\bel
&&\prod_{j=0}^{\infty} \prod_{m=-j \atop j-m=\text{even}}^{j}  \left(   (j +1-\Delta) \beta- \beta r D_{3} \right){}^{-1}  \non
&&=\prod_{n\in \Z} \prod_{j=0}^{\infty} \prod_{m=-j \atop j-m=\text{even}}^{j}  (-2\pi i n-i \beta r \rho(a)+(j+m+1) \beta_1+(j+1-\Delta-m) \beta_2 -F_l M_l){}^{-1}.  \non
\ee
Next we evaluate the one-loop determinant of the fermions.
We expand the $\psi$ by the spinor harmonics  $\chi^{s}_{j m}$ as
\bel
&&\psi=\sum_{\rho} \sum_{s=\pm} \sum_{j = \frac{1}{2}}^{\infty} \mathop{{\sum}'}_{m=-j}^{j}  
\psi^{ \rho, s}_{j m} (\tau) \chi^{s}_{j m} (\mth, \vphi ) E_{\rho}. 
\label{Fermidash}
\ee
Here the sum $\mathop{{\sum}'}$ runs over the following modes under the boundary condition (\ref{Ncondition}):
\bel
&& \chi^{+}_{j m}: j-m=\text{odd}, \quad  \chi^{-}_{j m}: j-m=\text{even}.
\ee
Then the action of the chiral multiplet (\ref{Lagchiral}) can be expanded at the quadratic order
\bel
S^{(2)}_{\text{chi}} \Big|_{\text{fer}}  
&&=\frac{i}{2r}\int_{S^1}\sum_{\rho}  \sum_{j}   \sum_{m; j-m=\text{even}} 
 \bpsi^{ \rho, +}_{j,-m}   \left(  j  -   r D_{3} +  1-\Delta     \right)
 \psi^{\rho, -}_{j, m} \cdot  (-1)^{m-1/2} \non
&& +  \frac{i}{2r}\int_{S^1}\sum_{\rho} \sum_{j}  \sum_{m; j-m=\text{odd}}
 \bpsi^{ \rho, -}_{j,-m}   \left(  j  +   r D_{3} +  \Delta    \right)
 \psi^{\rho, +}_{j, m} \cdot  (-1)^{m-1/2}  . 
\ee
The factor $\left(  j  -   r D_{3} +  1-\Delta     \right)$ contributes 
to the one-loop determinant of the chiral multiplet with $j=j'+\frac{1}{2}$, $m=m'+\frac{1}{2}$:
\bel
&&\prod_{j=\frac{1}{2}}^{\infty} 
 \text{Det}\left(   (j -   \Delta +  1) \beta- \beta r D_{3} \right)  \non
&&=\prod_{n \in \Z} \prod_{j'=0}^{\infty} \prod_{m'=-j' \atop j'-m'=\text{even}}^{j'} 
\prod_{l} (-2\pi i n-i \beta r \rho(a)+(j'+m'+1) \beta_1+(j'+1-\Delta-m') \beta_2 -F_l M_l). \non
\ee
The other factor $\left(  j  +   r D_{3} +\Delta     \right)$ contributes to the 
one-loop determinant of the chiral multiplet with $j=j'-\frac{1}{2}$, $m=m'+\frac{1}{2}$:
\bel
&&\prod_{j=\frac{1}{2}}^{\infty}  
\text{Det} \left(   ( j   +\Delta) \beta+ \beta r D_{3} \right)  \non
&&=\prod_{n \in \Z} \prod_{j'=1}^{\infty} \prod_{m'=-j' \atop j'-m'=\text{odd} }^{j'-2} \prod_{l} 
(2\pi i n+i  \beta r \rho(a)+(j'-m') \beta_1+(j'+\Delta +m') \beta_2 +F_l M_l).\non
\ee
Then the one-loop determinant of the chiral multiplet with the Neumann boundary condition is given by 
\bel
Z^{\text{3d.chi}.\text{N}}_{1\text{-loop}}&&
=\prod_{\rho} \prod_{n \in \mathbb{Z}} \prod_{j=0}^{\infty}   \prod_{l} (2\pi i n+i \beta r \rho(a)+(2j+\Delta) \beta_2 +F_l M_l)^{-1} \non
&&=\prod_{\rho}    \prod_{l} {e^{ \mathcal{E} (i\beta r \rho(a)+\Delta \beta_2+F_l M_l)}} ( e^{-i \beta r \rho(a) -F_l M_l} q^{\Delta} ; q^{2} )^{-1}.
\ee
The one-loop determinant of the chiral multiplet also does not depend on the parameter $\beta_1$.

\subsubsection{Dirichlet boundary condition}
Next we  evaluate the bosonic one-loop determinant for the chiral multiplet with the Dirichlet boundary condition (\ref{Dcondition}).
At the quadratic order, the bosonic part of the action is expanded as
 \bel
&&{S}^{(2)}_{\text{chi}} \Big|_{\text{bos}} 
=\frac{1}{2 r^2} \int_{S^1}
\sum_{j=0}^{\infty} \sum_{m=-j \atop j-m=\text{odd}}^{j} \sum_{\rho} \bar{\phi}^{\rho}_{ j, m}  
\left( j +\Delta +  r D_{3} \right) \left(  j+1 -\Delta - r D_3 \right)   \phi^{\rho}_{j m}. \non
\label{chiralDmode}
\ee
The factor $ \left(   j +\Delta+r D_{3} \right)$ in (\ref{chiralDmode}) contributes to  
the  one-loop determinant of the bosonic fields  as
\bel
&&\prod_{j=1}^{\infty} \prod_{m=-j+1 \atop j-m=\text{odd}}^{j-1}  
\mathrm{Det} \left(   (j +\Delta) \beta+ \beta r D_{3} \right){}^{-1}  \non
&&=\prod_{n \in \Z} \prod_{j=1}^{\infty} \prod_{m=-j+1 \atop j-m=\text{odd}}^{j-1} \prod_{l} 
 (2\pi i n+i \beta r \rho(a)+(j-m) \beta_1+(j+\Delta+m) \beta_2 +F_l M_l){}^{-1}. \nonumber
\ee
The other factor $\left(   j+1 -\Delta - r D_3 \right)$ contributes to 
the   one-loop determinant of the bosonic fields as 
\bel
&&\prod_{j=1}^{\infty} \prod_{m=-j+1 \atop j-m=\text{odd}}^{j-1} \prod_{l} 
\left(   (j +1-\Delta) \beta- \beta r D_{3} \right){}^{-1}  \non
&&=\prod_{n\in \Z} \prod_{j=1}^{\infty} \prod_{m=-j+1 \atop j-m=\text{odd}}^{j-1}  (-2\pi i n-i \beta r \rho(a)+(j+m+1) \beta_1+(j+1-\Delta-m) \beta_2 -F_l M_l){}^{-1}.  \nonumber
\ee
Next we evaluate the one-loop determinant of the fermions.
In this case, the sum $\mathop{{\sum}'}$ in (\ref{Fermidash}) runs over the following  modes under the boundary condition (\ref{Dcondition}):
\bel
&& \chi^{+}_{j m}: j-m=\text{even}, \quad 
  \chi^{-}_{j m}: j-m=\text{odd}.
\ee
The action of the chiral multiplet (\ref{Lagchiral}) can be expanded at the quadratic order 
\bel
S^{(2)}_{\text{chi}} \Big|_{\text{fer}} 
 &&
=\frac{i}{2r}\int_{S^1}\sum_{\rho}  \sum_{j}  \sum_{m; j-m=\text{odd}} 
   \bpsi^{ \rho, +}_{j,-m}   \left(  j  -   r D_{3} +  1-\Delta     \right)
 \psi^{\rho, -}_{j, m}  \cdot (-1)^{m-1/2} \non
&& + \frac{i}{2r} \int_{S^1}\sum_{\rho} \sum_{j}   \sum_{m; j-m=\text{even}}
 \bpsi^{ \rho, -}_{j,-m}   \left(  j  +   r D_{3} + \Delta    \right)
 \psi^{\rho, +}_{j, m}  \cdot (-1)^{m-1/2} .  \nonumber
\ee
Then, the factor $\left(  j  -   r D_{3} +  1-\Delta     \right)$ contributes 
to the one-loop determinant of the chiral multiplet 
\bel
&&\prod_{j=\frac{1}{2}}^{\infty} 
 \text{Det}\left(   (j -   \Delta +  1) \beta- \beta r D_{3} \right)  \non
&&=\prod_{n \in \Z} \prod_{j'=0}^{\infty} \prod_{m'=-j' -1\atop j'-m'=\text{odd} }^{j'-1} \prod_{l} (-2\pi i n-i \beta r \rho(a)+(j'+m'+1) \beta_1+(j'+1-\Delta-m') \beta_2 -F_l M_l), \nonumber
\ee
where we defined $j'=j-\frac{1}{2}, m'=m-\frac{1}{2}$. 
The other factor $\left(  j  +   r D_{3} +\Delta     \right)$ contributes to the  one-loop determinant of the 
chiral multiplet as
\bel
&&\prod_{j=\frac{1}{2}}^{\infty}  
\text{Det} \left(   ( j   +\Delta) \beta+ \beta r D_{3} \right)  \non
&&=\prod_{n \in \Z} \prod_{j'=1}^{\infty} \prod_{m'=-j' +1\atop j'-m'=\text{odd}}^{j'-1} \prod_{l} (2\pi i n+i  \beta r \rho(a)+(j'-m') \beta_1+(j'+\Delta+m') \beta_2 +F_l M_l).\nonumber
\ee
Here we defined $j'=j+\frac{1}{2}, m':=m-\frac{1}{2}$. 
Thus the one-loop determinant of the chiral multiplet with the Dirichlet boundary condition is given by 
\bel
Z^{\text{3d.chi}.\text{D}}_{1\text{-loop}}&&
=\prod_{\rho } \prod_{n \in \mathbb{Z}} \prod_{j=0}^{\infty}   \prod_{l} (-2\pi i n-i \beta r \rho(a)+(2j+2-\Delta) \beta_2 -F_l M_l) \non
&&=\prod_{\rho }  \prod_{l} {e^{- \mathcal{E} (-i\beta r \rho(a)+(2-\Delta) \beta_2-F_l M_l)}} ( e^{i \beta r \rho(a) -F_l M_l} q^{2-\Delta} ; q^{2} )_{\infty}.
\ee

\subsection{2d $\mathcal{N}=(0,2)$ chiral multiplet}
We have  boundary  theories with the $\mathcal{N}=(0,2)$ supersymmetry. 
The Lagrangian for the $\mathcal{N}=(0,2)$ chiral multiplet is given by  
\bel
 \mathcal{L}^{\mathcal{N}=(0,2)}_{\text{chi}}
&&=   \bphi (D_{\hat{2}}-iD_{\hat{3}}) (D_{\hat{2}}+iD_{\hat{3}}) \phi+\frac{1}{2} \bpsi ' (D_{\hat{2}}-iD_{\hat{3}})  \psi '
\non
&& \qquad  \quad +\frac{ i\Delta }{r} \bphi (D_{\hat{2}}-iD_{\hat{3}})  \phi   +  
i\bphi \blam_1 \psi ' + i \bpsi_1 \lam_1 \phi + \bphi (F_{\hat{2} \hat{3}} -i\hat{D}) \phi  .
\ee
The one-loop determinant of the bosonic part is given by
\bel
&&\mathrm{Det} (D_{\hat{2}}+iD_{\hat{3}}+\frac{i\Delta}{r})(D_{\hat{2}}-iD_{\hat{3}}) \non
&&\cong 
\prod_{\rho}  \prod_{l} \prod_{m \in \Z}  \prod_{n \in \Z} ( 2\pi i n+i\beta r \rho(a)+(2m+\Delta) \beta_2+F_l M_l ) \non
&& \quad \times 
\prod_{\rho}  \prod_{l} \prod_{m \in \Z} \prod_{n \in \Z} (2\pi i n+i\beta r \rho(a)-(\Delta+2m)\beta_1+F_l M_l ) .
\ee
The one-loop determinant of the fermionic part is given by
\bel
\mathrm{Det} (D_{\hat{2}}- iD_{\hat{3}})&&\cong  
\prod_{\rho} \prod_{m' \in \Z } \prod_{n \in \Z} 
(  2\pi i n+ i \beta \rho(a)- (\Delta +2m')\beta_1+F_l M_l) .
\ee
Then the one-loop determinant of the chiral multiplet can be written by 
\bel
Z^{\text{2d.chi}}_{1\text{-loop}}=\prod_{\rho } \prod_{l } 
e^{2\mathcal{E}( i\beta r \rho(a)+\Delta \beta_2+F_l M_l)}
{\th(e^{- i \beta r \rho(a)- F_l M_l } q^{\Delta}; q^2 )}^{-1}.
\ee

\subsection{2d $\mathcal{N}=(0,2)$ Fermi multiplet}
Next we evaluate the  one-loop determinant of the Fermi multiplet  on the torus $T^2$:
\bel
 \mathcal{L}^{\mathcal{N}=(0,2)}_{\text{Fermi}}
=     -  \bar{\Psi} ( D_{\hat{2}}+i D_{\hat{3}} ) \Psi + 2\bar{G} G + 2\bar{E} E -  \bpsi_{E} \Psi-\bar{\Psi} \psi_{E}
 + \frac{i}{r} (  1 - \tilde{\Delta} ) \bar{\Psi} \Psi.
  \nonumber
\ee  
At the quadratic order, the one-loop contribution comes from the determinant of the fermion: 
\bel
Z^{\text{2d.Fermi}}_{1\text{-loop}}&&= \mathrm{Det}
 \Bigl[   -(  D_{\hat{2}} +iD_{\hat{3}} )+\frac{i}{r} (  1 - \tilde{\Delta} )   \Bigr]\non
&&\cong 
\prod_{a} \prod_{\rho}\prod_{m' \in \Z} \prod_{n \in \Z} (2\pi i n+(2m'+\tilde{\Delta} )\beta_2
+i \beta r \rho(a)+F_a M_a)  \non
&&=\prod_{a} \prod_{\rho} 
e^{-2\mathcal{E}(i\beta r \rho(a)+\tilde{\Delta} \beta_2+F_a M_a)} 
\th(e^{-i\beta r \rho(a)  -F_a M_a } q^{\tilde{\Delta}};q^2) .
\ee

\end{document}